\documentclass[final]{elsarticle}
\usepackage{hyperref}

\usepackage{rotating}
\usepackage{color}
\usepackage{amsmath}
\usepackage{amssymb}
\usepackage{amsthm}
\usepackage{bbm}
\usepackage{booktabs}
\usepackage{multicol}
\usepackage{caption}
\usepackage{subcaption}
\usepackage{multirow}
\usepackage{stackrel}
\usepackage{hhline}
\usepackage{makecell}
\usepackage{nicefrac}
\usepackage[margin=1in]{geometry}
\usepackage{csquotes}

\newcommand{\E}[1]{\mathbb{E}\left[ #1\right]}
\newcommand{\Var}[1]{\mathbb{V}\mathrm{ar}\left( #1\right)}
\newcommand{\indV}[1]{\mathbbm{1}_{\left\{ #1 \right\}}} 
\newcommand{\PhiDifNull}[2]{\left( \Phi\left({#1}\right) - \Phi\left({#2}\right)\right)}
\newcommand{\PhiDifFin}[3]{\left( \Phi\left({#1} #3\gamma\sqrt{T}\right) - \Phi\left({#2} #3 \gamma\sqrt{T}\right)\right)}
\newcommand{\ExpTwoPar}[2]{\exp\left({#1}\left( r+0.5\gamma^2\right)T + {#2}\gamma^2T\right)} 

\newcommand{\Q}{\mathbb{Q}}
\newcommand{\R}{\mathbb{R}}

\newcommand{\tin}{t \in [0, T]}

\DeclareMathOperator*{\argmax}{argmax}

\allowdisplaybreaks

\theoremstyle{definition}
\newtheorem{assumption}{Assumption}
\newtheorem{theorem}[assumption]{Theorem}
\newtheorem{lemma}[assumption]{Lemma}
\newtheorem{corollary}[assumption]{Corollary}
\newtheorem{proposition}[assumption]{Proposition}
\newtheorem{definition}[assumption]{Definition}
\newtheorem*{remark}{Remark}

\newtheorem{taggedlemmax}{Lemma}
\newenvironment{taggedlemma}[1]
 {\renewcommand\thetaggedlemmax{#1}\taggedlemmax}
 {\endtaggedlemmax}
 
 \newtheorem{taggedpropx}{Proposition}
\newenvironment{taggedprop}[1]
 {\renewcommand\thetaggedpropx{#1}\taggedpropx}
 {\endtaggedpropx}

\journal{Journal of Banking and Finance}

\biboptions{authoryear}

\begin{document}

\begin{frontmatter}

\title{Optimal fees in hedge funds with first-loss compensation} 
\author[escobar_address]{M. Escobar-Anel\corref{cor1}}
\ead{marcos.escobar@uwo.ca}

\author[havrylenko_zagst_address]{Y. Havrylenko}
\cortext[cor1]{Corresponding author}
\ead{yevhen.havrylenko@tum.de}

\author[havrylenko_zagst_address]{R. Zagst}
\ead{zagst@tum.de}

\address[escobar_address]{Department of Statistical and Actuarial Sciences, Western University, 1151 Richmond street, London, Ontario, Canada}
\address[havrylenko_zagst_address]{Chair of Mathematical Finance, Technical University of Munich, Parkring 11, 85748 Garching bei M\"unchen, Germany}

\begin{abstract}

Hedge fund managers with the first-loss scheme charge a management fee, a performance fee and guarantee to cover a certain amount of investors' potential losses. We study how parties can choose a mutually preferred first-loss scheme in a hedge fund with the manager’s first-loss deposit and investors’ assets segregated. For that, we solve the manager’s non-concave utility maximization problem, calculate Pareto optimal first-loss schemes and maximize a decision criterion on this set. The traditional $2\%$ management and $20\%$ performance fees are found to be not Pareto optimal, neither are common first-loss fee arrangements. The preferred first-loss coverage guarantee is increasing as the investor’s risk-aversion or the interest rate increases. It decreases as the manager’s risk-aversion or the market price of risk increases. The more risk averse the investor or the higher the interest rate, the larger is the preferred performance fee. The preferred fee schemes significantly decrease the fund's volatility.

\end{abstract}


\end{frontmatter}

    
\section{Introduction}
\textbf{Motivation.} A hedge fund is an asset managing company that oversees pooled investment vehicles and whose clients are solely accredited investors. It faces less regulation than pension funds and mutual funds. Its investment strategies usually exploit rare opportunities and the key determinant of its performance are the skills of the hedge fund managers. Therefore, the manager's compensation usually has a management fee proportional to the assets under management (AUM) and a performance fee depending on the hedge fund's performance.
The size of these fees often follows the so-called \enquote{$2\,\&\, 20$} rule, i.e. $2\%$ of the AUM is the management fee and $20\%$ of the fund's profit above a set benchmark is the performance fee.

In recent years, the hedge fund sector has faced a lot of criticism of its high fees and lukewarm performance. According to the Financial Times\footnote{See \url{https://www.ft.com/content/b8ca99da-9782-11e7-a652-cde3f882dd7b}}, investors pulled billions of dollars from hedge funds in 2016 and resorted to passive or private equity strategies.
To regain investors' trust and make investments in hedge funds more attractive, some hedge funds started offering a first-loss compensation scheme\footnote{See the article about the first-loss capital in the USA at \url{https://www.marketwatch.com/story/more-hedge-funds-lured-to-new-source-of-capital-2011-05-23}}. In this fee structure, the hedge fund manager guarantees to cover with her own money potential losses up to a certain percentage of the investor's initial endowment. As compensation for this first-loss coverage guarantee, the manager may charge higher management and/or performance fees. The natural questions that arise are about the trade-off between the parameters of first-loss scheme and what a \enquote{fair} or \enquote{well-balanced} first-loss fee structure should be.

\textbf{Aim.} The research question we answer  is which first-loss schemes can be seen as fair and optimal for both parties, i.e. which management fee, performance fee and first-loss coverage guarantee are mutually preferred by both, the manager and the investor. Within standard economics paradigms of rational expectations and utility maximization, we also study the impact of parameters of the financial market and of decision makers' risk aversion on such mutually preferred fee arrangements.  Our research object is the first-loss fee structure and not the related investment strategy. The latter can be determined by replicating the fund's optimal terminal value derived in this paper.

\textbf{Contribution.}
To the best of our knowledge, we are first to analyze first-loss compensation schemes based on the criterion of Pareto optimality and to investigate how hedge fund managers and investors can reasonably select a single Pareto optimal first-loss fee structure. Conducting extensive numerical studies for hyperbolic absolute risk aversion (HARA) utility functions with parameter values consistent with current risk appetites in the hedge-fund sector, we find that the common $2\%$ management and $20\%$ performance fees are not Pareto optimal in the traditional scheme. The \enquote{closest}  Pareto optimal fee structure in the traditional setting has a $0\%$ management fee and a $20\%$ performance fee,  which may explain the current trend of decreasing management fees in hedge funds that use traditional managerial compensation.  Furthermore, the first-loss fee structures typically used by hedge funds (with a performance fee around $40\%$ and a first-loss coverage guarantee around $10\%$, see \cite{Djerroud2016}, \cite{He2018}) are not Pareto optimal either.
The Pareto optimal first-loss fee structure that maximizes the hedge-fund's Sharpe ratio has a management fee of $5\%$, a performance fee about $35\%$ and a first-loss coverage guarantee around $25\%$. However, the manager might not agree on this fee structure as her expected utility is lower than the one for the $(0\%, 20\%)$ traditional fee structure. Using the same criterion for the first-loss fee structure selection but requiring that both parties are better off when the traditional fee structure is replaced by the first-loss one, the decision makers should agree on a management fee of $5\%$, a performance fee about $48\%$ and a first-loss coverage guarantee around $24\%$. The methodology we use yields a preferred Pareto optimal first-loss fee structure that is fair to both parties and  decreases significantly the hedge fund’s risk in comparison to the traditional fee structure. Also to the best of our knowledge, we are the first to study the trade-off between the parameters of the first-loss scheme in the expected utility framework, which can be of great help in the fee structure negotiation process.

\textbf{Related literature.}
To solve the manager's non-concave utility maximization problem, where non-concavity arises due to her payoff profile, we combine the martingale approach (\cite{Karatzas1987}, \cite{Cox1989}) and the concavification technique. The latter technique is based on the construction of the concave envelope of a utility function. It dates back to \cite{Aumann1965}. This technique was first used in the context of managerial compensation in \cite{Carpenter2000}. Later this result was extended to more general managerial compensation schemes (\cite{Larsen2005}, \cite{Bichuch2014}). \cite{Reichlin2013} proves the existence and several fundamental properties of the solution to unconstrained portfolio optimization problems in a very general framework. However, the author does not provide any specific payoffs. We take advantage of our specific setting to derive explicitly the manager's optimal terminal wealth without the need for convex conjugates and sub-differentials.

The literature on managerial compensation in hedge funds mainly focuses on the traditional fee structure, in particular on the link between risk taking and performance fees. Almost no papers can be found which search for an \enquote{equilibrium} fee structure.  \cite{Carpenter2000} analyzes the impact of performance fees on the optimal investment strategy of a manager holding an option on the fund's assets and having preferences modeled by a HARA utility function. The author finds that the performance fee causes the manager to reduce the fund's risk.  \cite{Kouwenberg2007} analyze how the performance fee and the manager's own investment in the fund influence the risk aversion of a manager whose preferences are modeled using prospect theory. In contrast to \cite{Carpenter2000} the researchers find that performance fees increase the manager's risk appetite. In their broad empirical study, they find though, that for individual hedge funds there is no significant relation between volatility and performance fees. \cite{Hodder2007} consider a hedge fund manager with a power utility and a traditional compensation scheme. They come to the conclusion that although the manager's risk-taking may change drastically depending on the fund value within a one-year investment period, this effect is moderate over longer investment horizons. \cite{Guasoni2016} study hedge funds with traditional compensation schemes where performance fees are high-water marked. The researchers find that such performance fees increase risk-taking for managers with typical levels of risk aversion. \cite{Zou2017} analyzes traditional fee structures and derives optimal investment strategies of a hedge fund manager with a piecewise exponential utility and an S-shaped utility from the Cumulative Prospect Theory (CPT) framework. The researcher concludes that the manager pursues less risky investment strategies when her loss aversion, risk aversion, ownership in the fund or the management fee ratio increases. When the performance fee increases, though, the manager acts riskier.  \cite{EscobarAnel2018} is the only paper that studies the question of how the manager and the investor can agree on a single mutually preferred traditional fee structure. They propose two procedures on how to select a fair traditional fee structure for both parties and conclude that for reasonable market parameters the fee arrangement with $0.5\%$ management and $30.7\%$ performance fee stands out as a fair one. We apply their approach based on Pareto optimality and Sharpe ratio maximization in the context of first-loss schemes. The fee structure $(0.5\%, 30.7\%, 0\%)$ is not Pareto optimal in the presence of first-loss coverage. It is  almost twice more volatile and yields to a Sharpe ratio of that fund that is about $20\%$ lower in comparison to the preferred first-loss fee structure.

 There are two papers that analyze the novel first-loss fee structure, although from different angles.  \cite{Djerroud2016} examine the first-loss scheme in the derivative pricing framework, but do not optimize portfolios. They conduct a cost-benefit analysis of particular fee structures and calculate \enquote{fair} performance fees, where the investor has a payof{}f with present value equal to her initial cash injection. However, they do not search for an optimal or an \enquote{equilibrium} fee structure. \cite{He2018} consider a hedge fund with a manager whose capital is invested in the fund, i.e. the manager's and the investor's money is commingled. The authors refer to the proportion of the fund that belongs to the manager as the managerial ownership ratio and consider the management fee to be part of it. Working in the CPT framework, they conclude that the first-loss scheme with $30\%$ performance fee and $10\%$ managerial ownership ratio used to cover first loss is better for both the investor and the manager than the fee structure with $20\%$ performance fee and $10\%$ managerial ownership ratio that is not used to cover any potential losses. However, the researchers do not investigate if the suggested first-loss fee structure $(0\%, 30\%, 10\%)$ is Pareto optimal for representative managers and investors. As opposed to \cite{He2018}, we consider the investors' assets being segregated from the  managers' money and follow the goal of finding Pareto optimal fee structures.

\textbf{Outline.} The remainder of the paper is organized as follows. In Section \ref{sec:model_setup} we introduce the financial market model as well as the model of a hedge fund with first-loss compensation. In Section \ref{sec:problem_setting} we state the portfolio optimization problem of the hedge fund manager and the problem of preferred fee structure selection. Furthermore, we provide theoretical results necessary for solving the former problem and present a rational methodology for determining a single first-loss fee that both parties may agree on. Considering the manager and the investor equipped with HARA utility functions, we solve the manager's optimization problem in Section \ref{sec:hara_theoretical_results} and derive the value function of each party. In Section \ref{sec:hara_numerical_studies} we select reasonable model parameters and conduct numerical studies. Finally, Section \ref{sec:conclusion} concludes. The proofs of the main results are provided in  \ref{app:proofs}. \ref{app:aux_proofs} contains supplementary results.
\section{Problem setting}\label{sec:model_setup}
\subsection{Financial market model}
Let $T>0$ be a finite time horizon and $(W_{t})_{\tin}$ be a one-dimensional Wiener process on a filtered probability space $\left( \Omega, \mathcal{F}, \left(  \mathcal{F}_t\right)_{\tin}, \mathbb{Q}\right)$ that satisfies usual conditions of completeness and right-continuity.
Consider the Black-Scholes financial market model with one risk-free asset and one risky asset, whose price dynamics under the real-world probability measure $\Q$ are given, respectively, by
\begin{equation*}
	\begin{aligned}
 		dS_{0, t} &= rS_{0, t}dt,\,\,\, S_{0,0} = 1;\\
		dS_{1, t} & = \mu S_{1, t} dt + \sigma S_{1, t} dW_{t},\,\,\, S_{1, 0} = s_1 > 0,
	\end{aligned}
\end{equation*}
where $r$ is a constant interest rate, $\mu$ is a constant drift and $\sigma >0$ is a constant volatility, and $s_1$ is the initial price of the risky asset.

In the following, we denote the market price of risk by $\gamma:=\sigma^{-1}(\mu - r)$ and the state price density process by:
\begin{equation}\label{eq:state_price_density}
\tilde{Z}_{t} =  \exp \left( -\left(r +  \frac{1}{2} \gamma^2 \right)t - \gamma W_{t} \right),\,\,\, t \in [0, T].
\end{equation}

\subsection{Hedge-fund model}
We consider the hedge fund model presented in \cite{Djerroud2016}. There are two parties -- the hedge-fund manager and the investor. At time $0$ the investor entrusts her initial capital $I_0 > 0$ to the hedge fund, so that the initial value of the fund equals  $V_0 = I_0 := v_0$. The hedge-fund manager invests the investor's capital and manages the money until the end of the investment period. We denote the self-financing relative portfolio process by $\pi = (\pi_{0, t}, \pi_{1, t})_{\tin}'$. $\pi_{0, t}$ and $\pi_{1, t}$ are the fractions of the hedge-fund's capital invested in the risk-free asset and the risky asset respectively at time $t$, $\tin$. Then the portfolio value $V_{t}$, $\tin$, evolves as follows:
\begin{equation*}\label{eq:wealth_dynamics}
 dV_{t} = r V_{t} dt + \pi_{1, t}V_{t}((\mu - r)dt + \sigma dW_{t}),\,\,\, V_{0}=v_0.
\end{equation*}
Let $\Lambda(v_0)$ be the set of all investment strategies for which $V_{0} = v_0$ and $V_{t}\geq 0$ for all $\tin$.

Denote by $V_T$ the fund's terminal value. At the end of the investment period, the fund's  value is split between the manager and the investor:
\begin{equation*}
V_T= I(V_T) + M(V_T),
\end{equation*}
where $M(V_T)$ is the terminal wealth of the manager and $I(V_T)$ is the terminal wealth of the investor.

The manager's payof{}f is determined by the compensation scheme, i.e. the fee structure. In the first-loss compensation model, the manager
\begin{itemize}
\item \textit{charges} a management fee;
\item \textit{charges} a performance fee on the investor's net profit, if it is positive;
\item \textit{guarantees} to cover incurred loss up to a certain percentage of the initial capital.
\end{itemize}

There are two basic types of first-loss arrangements:
\begin{enumerate}
	\item the investor's assets and the manager's deposit are commingled; 
	\item the investor's assets and the manager's deposit account are segregated.
\end{enumerate}
The first type is usually better for the manager, as it gives the manager shareholder rights. The second type is preferred by investors, as it removes shareholder rights from managers. There are many other considerations dealing with how the deposit account is securitized, they are usually variants of these two basic ones. In this paper we analyze the second case, using the model of the first-loss fee structure from \cite{Djerroud2016}.


We assume a single payment at the end of a fixed term $T$, which also implies that the fees are not invested. Later in the case study, we set $T = 1$, which means that the fees are charged at the end of the investment year. The assumption that the fees are charged once at time $T$ is common in the literature (see e.g. \cite{Kouwenberg2007},  \cite{He2018}, \cite{Zou2017}, \cite{EscobarAnel2018}).

We denote by $m \in [0, 1]$ the fixed share of $v_0$ that is charged by the hedge fund manager at time $t=T$. We call both $m$ and $m v_0$ the management fee. It will be clear from the context which term is meant.

In case the wealth generated for the investor is greater than the initial endowment, i.e. $V_T- m v_0 > v_0$,  the manager charges the share $\alpha\in (0, 1]$ of the capital surplus\footnote{We exclude $\alpha = 0$ for several reasons. First, the performance fee is a distinct feature of the hedge fund industry. Second, $\alpha = 0$ requires special treatment in the derivation of the fund's optimal terminal value, as in this case the manager's utility function is flat for $v \geq (1+m)v_0$. This would make the paper longer without contributing much to the aim of our paper. Third, we need $\alpha > 0$ for proving the existence of first best Pareto optimal fee structures, see Proposition \ref{prop:nonlin_opt_solution_existence} in Subsection \ref{subsec:general_setting}}. We refer to both $\alpha$ and $\alpha (V_T - m v_0 - v_0)^{+}$ as the performance fee\footnote{$x^+ = \max\{x, 0\}$ for $x \in \R$}. It will be clear from the context which element is meant.

We denote by $c\in [0, 1]$ the maximal share of $v_0$ paid by the manager to the investor if the latter faces a loss at time $T$. The loss occurs whenever the terminal portfolio value less the management fee is lower than the investor's initial investment. We refer to both $c$ and $c v_0$ as the manager's first-loss coverage guarantee.

Then the investor's terminal wealth in the first-loss  scheme is given by:
\begin{equation}\label{eq:terminal_wealth_investor}
I(V_T) = \left\{
\begin{aligned}
	&V_T + v_0(c - m),&&\text{if } V_T - mv_0 < (1 - c) v_0;\\
	&v_0,\,&& \text{if } (1 - c)v_0 \leq V_T -mv_0  < v_0;  \\
	&V_T - mv_0 - \alpha (V_T - (1+m)v_0),&&\text{if }V_T - mv_0 \geq v_0. \\
\end{aligned}
\right.
\end{equation}

Using (\ref{eq:terminal_wealth_investor}) and the relation $M(V_T) = V_T - I(V_T)$, we obtain the following terminal payof{}f of the manager:
\begin{equation}\label{eq:terminal_wealth_manager}
M(V_T) = \left\{
\begin{aligned}
	&v_0(m - c), & &  \text{if } V_T - mv_0 < (1 - c) v_0;\\
	&V_T - v_0, & & \text{if } (1 - c)v_0 \leq V_T -mv_0  < v_0; \\
	&m v_0 + \alpha (V_T  - (1+m)v_0), & &\text{if }V_T - mv_0 \geq v_0. \\
\end{aligned}
\right.
\end{equation}

\begin{figure}[!htbp]
\begin{subfigure}{.5\textwidth}
  \centering
  \includegraphics[width=\linewidth]{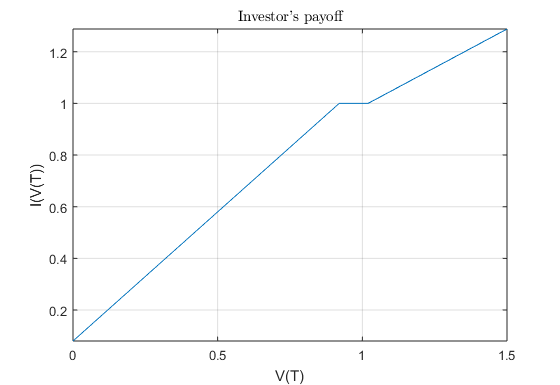}
  \caption{Investor's payoff}
\label{sfig:man_payoff}
\end{subfigure}%
\begin{subfigure}{.5\textwidth}
  \centering
  \includegraphics[width=\linewidth]{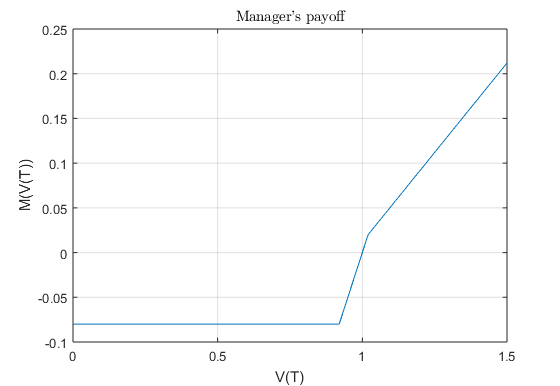}
  \caption{Manager's payoff}
  \label{sfig:inv_payoff}
\end{subfigure}
\caption{Parties' payoffs for $m = 2\%, \alpha = 40\%, c = 10\%$}
\label{fig:payoff_profiles}
\end{figure}

From (\ref{eq:terminal_wealth_investor}) and (\ref{eq:terminal_wealth_manager}) we see that the terminal wealth of each party is a continuous non-decreasing piecewise linear function of the fund's terminal value. Figure \ref{fig:payoff_profiles} illustrates the payoffs of both parties.

Note that the terminal wealth of the investor and that of the manager can attain negative values. For example, the manager's payof{}f may be negative when the fund's terminal value is sufficiently small and the first-loss guarantee offered by the manager is greater than the charged management fee. The investor's terminal wealth may be negative when the fund's terminal value is sufficiently small and the provided first-loss coverage is less than the management fee the investor paid.

Theoretically, each parameter of the first-loss fee structure can attain values between $0\%$ and $100\%$. A few hedge funds have experimented with negative management fees to attract clients. In Section \ref{sec:hara_numerical_studies}, where we present the results of our numerical studies, we consider $\mathcal{P} = \{ (m, \alpha, c) : m \in [0\%, 5\%], \alpha \in [0.1\%, 50\%], c \in [0\%, 30\%]\}$, which is motivated by practical considerations described later. We refer to this set as the set of admissible fee structures.

\subsection{Portfolio optimization and fee structure preferences} \label{sec:problem_setting} 
We assume that the manager's and the investor's preferences are determined by utility functions $\tilde{U}_{M}$ and $\tilde{U}_{I}$ respectively, which are strictly increasing, strictly concave, continuously differentiable and satisfy Inada conditions $\lim_{v \downarrow 0} \tilde{U}_{P}'(v) = +\infty$ and $\lim_{v \uparrow +\infty} \tilde{U}_{P}'(v) = 0$, $P \in \{ I, M \}$.
Since the manager's minimal terminal payof{}f equals $(m - c)v_0$ (for all $V_T \in [0, (1+m - c)v_0)$), we impose the condition $\tilde{U}_M((m - c)v_0) > -\infty$. Since the investor's minimal terminal payof{}f is $(c - m)v_0$, we impose the condition $\tilde{U}_I((c - m)v_0) > -\infty$. For instance, these inequalities are fulfilled for HARA utility functions or exponential utility functions with suitably chosen parameters.

We denote the parties' utility functions as functions of the fund's terminal value by
\begin{equation*}
		U_M(V_T) := \tilde{U}_M(M(V_T)), \qquad U_I(V_T) := \tilde{U}_I(I(V_T)).
\end{equation*}


Since we do not want to overcomplicate the model and focus on the fair fee selection process, we do not model the investments of the manager and of the investor outside of the hedge fund. In general, 
the manager solves the following portfolio optimization problem after fixing her first-loss fee structure:
\begin{equation*}
\begin{aligned}
 \max_{\pi}\, \, &  \mathbb{E}\left[{U}_M(V_T) \right]; \\
 \text{s.t.}\,\,\,\, &  \pi \in \Lambda(v_0)\text{ and } \E{{U}_M(V_T)^{-}} < +\infty.
\end{aligned}
\end{equation*}

As we focus on the fee structure selection in this paper, instead of the manager's full portfolio optimization problem, it suffices to consider only the related terminal portfolio value problem\footnote{This is possible due to the martingale approach to portfolio selection problems (see e.g. \cite{Karatzas2016})}:
\begin{equation}\label{eq:nonc_optim_wealth_problem_man}\tag{$P_{M}$}
\begin{aligned}
\max_{V_T}\, & \mathbb{E}[U_M(V_T)]; \\
\text{s.t.} \,& \mathbb{E}[\tilde{Z}_TV_T] \leq v_0; \\
 & V_T\geq 0. \\
\end{aligned}
\end{equation}

Note that the only influence the investor has on the fund's terminal value is the (first-loss) fee structure the parties negotiate at time $0$. However, there are infinitely many admissible fees. The crucial question is how the manager and the investor can agree rationally on a single mutually preferred fee, such that the interests of both parties are taken into account.

We denote the fund's optimal terminal value by $V_T^*:=V_T^*(m, \alpha, c)$.  Obviously, various fee structures lead to different optimal terminal values of the fund and, hence, different payoffs to the parties'. Therefore, the way how the manager and the investor agree on the fee structure is of high importance. In our view, it should lead to a fee structure that does not favor one party over the other and has a positive impact on the fund's overall performance.

We denote the parties' expected utilities evaluated at $V_T^*$ by
\begin{equation*}
		\phi_M(m, \alpha, c) := \E{U_M(V_T^*)} \text{ and } \phi_I(m, \alpha, c) := \E{U_I(V_T^*)},
\end{equation*}
and refer to them as value functions.

We examine optimal first-loss fee structures $(m, \alpha, c) \in \mathcal{P}$ in an analogous way \cite{Filipovic2015} investigated optimal investment and premium policies.

\begin{definition}[First best Pareto optimal fee structures]\label{def:pareto_optimal_fee}
A fee structure $(m^*, \alpha^*, c^*)$ is first best Pareto optimal (FBPO) if it solves the optimization problem
  \begin{equation}\label{eq:pareto_optimization_problem}\tag{$P_{I|M}$}
    \begin{aligned}
      \max_{m, \alpha, c}\, & \phi_I(m, \alpha, c); \\
      s.t. \,& \phi_M(m, \alpha, c) \geq \phi_{min}; \\
      & (m, \alpha, c) \in \mathcal{P};
    \end{aligned}
  \end{equation}
  for some reservation utility level $\phi_{min} \in \mathbb{R}$ of the manager. When we need to emphasize the dependence of this problem on the parameter $\phi_{\min}$, we write \ref{eq:pareto_optimization_problem}($\phi_{\min}$).
\end{definition}

We denote the set of all first best Pareto optimal fee structures by $\mathcal{P}_{FBPO}$ and refer to the set of all pairs $(\phi_M(m^*, \alpha^*, c^*), \phi_I(m^*, \alpha^*, c^*))$ such that $(m^*, \alpha^*, c^*) \in \mathcal{P}_{FBPO}$ as the Pareto frontier.

There are many ways how the manager and the investor can agree on a single FBPO fee structure. In this paper we focus on the hedge-fund's Sharpe ratio maximization as the criterion in the fee selection process. The Sharpe ratio is defined as follows:
\begin{equation*}
		  SR^*(m, \alpha, c):=SR(V_T^*(m, \alpha, c)) = \frac{\E{R(V_T^*)} - r}{\sqrt{\Var{R(V_T^*)}}} = \frac{\E{V_T^*} - v_0(1 + r)}{\sqrt{\Var{V_T^*}}}, 
\end{equation*}
where $R(V_T^*)$ denotes the rate of return of the hedge fund under the optimal investment strategy:
\begin{equation*}
		 R(V_T^*) = \frac{V_T^* - v_0}{v_0}.
\end{equation*}

We use this criterion for several reasons: the Sharpe ratio is a popular performance measure that can be observed in the hedge-fund sector, it is not based on utility functions and does not favor exclusively any of the parties. So we consider the fee structure selection as the process of solving the following optimization problem:
\begin{equation}\label{eq:eq_fee_opt_problem}\tag{$P_{SR}$}
  \begin{aligned}
    \max_{m, \alpha, c} & \,\,\, SR^*(m, \alpha, c); \\
    \text{s.t.} &\,\,\, (m, \alpha, c) \in \mathcal{P}_{FBPO}.
  \end{aligned}
\end{equation}
We denote the solution of Problem (\ref{eq:eq_fee_opt_problem}) as $(\hat{m}, \hat{\alpha},\hat{c})$ and call it the preferred fee structure.

\section{Solution approach}\label{sec:fund_optimal_V}
In the first part of this section, we describe theoretical and numerical approaches to finding the preferred fee structure  in general. In the second part of this section, we provide specific formulas for the case when decision makers have HARA utility functions. All proofs can be found in the appendix.
\subsection{General setting}\label{subsec:general_setting}

Using (\ref{eq:terminal_wealth_manager}), we can write the manager's utility function in the following way:

\begin{equation}\label{eq:utility_fct_manager_fl3}
\begin{aligned}
U_M(V_T) &:= \tilde{U}_M(M(V_T)) =  U_{M,1}(V_T)\mathbbm{1}_{[0, \Theta_1)}(V_T) + U_{M,2}(V_T)\mathbbm{1}_{[\Theta_1, \Theta_2)}(V_T)  + U_{M,3}(V_T)\mathbbm{1}_{[\Theta_2, +\infty)}(V_T),
\end{aligned}
\end{equation}
where $\Theta_1 = (1 + m - c )v_0, \Theta_2 = (1+m)v_0$, $U_{M,1}(v)=\tilde{U}_M((m - c)v_0))$,  $U_{M,2}(v)=\tilde{U}_M(V_T - v_0)$, $U_{M,3}(v)=\tilde{U}_M(m v_0 + \alpha (V_T  - m v_0 - v_0))$.

We denote $U_M'(\bar v - )$ and $U_M'(\bar v + )$ the left- and right-hand derivatives of $U_M$ at $\bar v \in \R$, and $U_M(\bar v - )$ and $U_M(\bar v + )$ for left- and right-hand limits of $U_M$ at $\bar v \in \R$.  To avoid any ambiguity, we set $U_M(v) = -\infty$ for $v < 0$, $U_M(0) = U_M(0+)$.

Obviously, $U_M$ is not concave, whence standard optimization tools cannot be applied to solve Problem (\ref{eq:nonc_optim_wealth_problem_man}). Therefore, we use the concavification technique (\cite{Carpenter2000}, \cite{Reichlin2013}, \cite{He2018}). First, we construct the concave envelope of $U_M$ and solve the resulting concavified problem defined below. Second, we show that the corresponding fund's optimal terminal value is also the solution of Problem (\ref{eq:nonc_optim_wealth_problem_man}).

There are several ways the concave envelope of a function may be defined. 
We follow \cite{Reichlin2013}.
\begin{definition}\label{def:concave_envelope}
The concave envelope $u_M$ of $U_M$ is the smallest concave function  $u_M: \R \rightarrow \R \cup \{ -\infty\}$ such that $U_M(v) \leq u_M(v)$ for all $v \in \mathbb{R}$.
\end{definition}

The next lemma shows in a constructive way that the concave envelope of $U_M$
exists and is unique.

\begin{lemma}\label{lem:concavification_3_pieces}
Let $U_M$ be defined according to (\ref{eq:utility_fct_manager_fl3}).
Assume that $U_{M,2}''$ and $U_{M,3}''$ exist on $(\Theta_1, \Theta_2)$ and $(\Theta_2, +\infty)$ respectively. Then there exists a unique $\theta_1 \geq \Theta_1$ such that the function
\begin{equation}\label{eq:concave_envelope_general}
u_M(v) = \left\{
\begin{aligned}
&-\infty, && \text{ if } v < 0;\\
&U_M(0) + s(\theta_1)v,& &\text{ if } v \in [0, \theta_1); \\
&U_M(v), & &\text{ if } v \geq \theta_1.
\end{aligned}\right.
\end{equation}
is the concave envelope of $U_M$, where $s(\theta_1) = \frac{U_M(\theta_1) - U_M(0)}{\theta_1}$.
\end{lemma}

For denoting the original function $U_M$ and its change points ($\Theta_1, \Theta_2$) we use uppercase letters, whereas we use the same but lowercase letters for the concave envelope $u_M$ and its change points ($\theta_1, \theta_2$). We resort to such notation, since these functions coincide in some parts but we still need to differentiate between them. Using (\ref{eq:utility_fct_manager_fl3}) and (\ref{eq:concave_envelope_general}), we obtain that the concave envelope is a continuous concave function that be written as:
\begin{equation}\label{eq:utility_piecewise}
u_M(v) =  \left\{
\begin{aligned}
&-\infty,\text{ if } v < 0;\\
&u_{M,1}(v) \mathbbm{1}_{[\theta_0, \theta_1)}(v) + u_{M,2}(v) \mathbbm{1}_{[\theta_1, \theta_2)}(v) + u_{M,3}(v) \mathbbm{1}_{[\theta_2, \theta_3)}(v),\text{ if } v \geq 0,
\end{aligned}\right.
\end{equation}
where $0 = \theta_0  < \theta_1 \leq \theta_2 < \theta_3 = +\infty$,  $u_{M,1}$ is a strictly increasing linear function, $u_{M,i},\,i \in \{2, 3\},$ are strictly increasing concave functions.
In our setting, $u_M$ can have between two or three pieces. According to the proof of Lemma \ref{lem:concavification_3_pieces}, $\Theta_2 \leq \theta_1 = \theta_2$ when $u_M$ consists of two pieces, otherwise $\Theta_1 < \theta_1 < \Theta_2 = \theta_2$.

Consider now the concavified version of Problem (\ref{eq:nonc_optim_wealth_problem_man}):
\begin{equation}\label{eq:conc_optim_wealth_problem_man}\tag{$P_{M}^{conc}$}
\begin{aligned}
\max_{V_T}\, & \mathbb{E}[u_M(V_T)]; \\
\text{s.t.} \,& \mathbb{E}[\tilde{Z}_T V_T] \leq v_0; \\
 & V_T\geq 0. \\
\end{aligned}
\end{equation}

In the next theorem we show how to solve Problem \eqref{eq:nonc_optim_wealth_problem_man} via Problem (\ref{eq:conc_optim_wealth_problem_man}).

\begin{theorem}\label{th:optim_w_linear_part_1}
Let $v(y, \tilde z)$ be the solution to the following pointwise optimization problem for any fixed $y > 0, \tilde z > 0$:
\begin{equation}\label{eq:point_optim_problem}
 \max_{v \geq 0} \{u_M(v) - y\cdot \tilde{z}\cdot v \},
\end{equation}
where $u_M$ is defined in (\ref{eq:utility_piecewise}). If the following integrability condition holds
\begin{equation}\label{as:integrability_1}
h(y) := \mathbb{E}\left[\tilde{Z}_T \cdot v^*(y, \tilde{Z}_T)\right] < +\infty\,\,\, \forall y \in (0, +\infty),
\end{equation}
then:
\begin{enumerate}
	\item there exists a unique $y^* \in (0, +\infty)$ such that $h(y^*) = v_0$;
	\item $V_T^* = v^*(y^*,\tilde{Z}_T)$ is the $\mathbb{Q}$-a.s. unique optimal terminal value in the concavified Problem (\ref{eq:conc_optim_wealth_problem_man});
	\item $V_T^* = v^*(y^*,\tilde{Z}_T)$ is the $\mathbb{Q}$-a.s. unique optimal terminal value in the original Problem (\ref{eq:nonc_optim_wealth_problem_man}).
\end{enumerate}
\end{theorem}

For many utility functions $U_M$ common in the literature (e.g. HARA, exponential), the fund's optimal terminal value $V_T^*$ can be found explicitly. Moreover, $\phi_P(m, \alpha, c) = \E{U_P(V_T^*)}$ for $P \in \{M, I\}$ can be found in a semi-explicit form. The next step is to obtain the set of the first-best Pareto optimal fee structures by solving the non-linear optimization Problem \eqref{eq:pareto_optimization_problem}.  The solution to this optimization problem exists, as proven in the following result.
\begin{proposition}\label{prop:nonlin_opt_solution_existence}
For any $\displaystyle \phi_{min} \in \left[\min_{(m, \alpha, c) \in \mathcal{P}}\phi_M(m, \alpha, c),\,\max_{(m, \alpha, c) \in \mathcal{P}}\phi_M(m, \alpha, c)\right]$ there exists \\$(m^*, \alpha^*, c^*)$ solving \eqref{eq:pareto_optimization_problem}.
\end{proposition}

In Problem  \eqref{eq:pareto_optimization_problem}, the objective function $\phi_I(m, \alpha, c)$ and the constraint function $\phi_M(m, \alpha, c)$ are in general quite complex and do not exhibit properties enabling us to solve this problem in closed form. Therefore, we will resort to numerical techniques, in particular, the Sequential Quadratic Programming (SQP) approach. For more information on this optimization technique, the interested reader is referred to \cite{Nocedal2006}.

\subsection{Solution for HARA Utility Functions}\label{sec:hara_theoretical_results}

In this section, we derive the fund's optimal terminal value as well as the parties' expected utility functions at the fund's optimal terminal value using the methodology from the previous section and assuming that the involved decision makers have HARA utility functions. 
Let $\tilde{U}_M$ be given by:
\begin{equation*}
\tilde{U}_M(v) = \frac{1}{1 - b_M} \left( v + a_M \right)^{1 - b_M},\,\, \tilde{U}_M: (-a_M, +\infty) \rightarrow \mathbb{R},
\end{equation*}
with $a_M \geq v_0(c - m)$ if $b_M \in (0, 1)$ and $a_M > v_0(c - m)$ if $b_M \in (1, +\infty)$.
In this case, we have the following concretization of  (\ref{eq:utility_fct_manager_fl3}):
\begin{equation}\label{eq:utility_hara_manager_fl3}
\begin{split}
U_M(V_T) &=  \tilde{U}_M(M(V_T)) =  \underbrace{\frac{1}{1-b_M}\left(v_0(m - c) + a_M\right)^{1-b_M}}_{U_{M,1}(V_T)}\cdot \mathbbm{1}_{[0,(1 + m - c )v_0)}(V_T) \\
& \,\,\,\,\,+\underbrace{ \frac{1}{1 - b_M}\left( V_T - v_0 + a_M \right)^{1-b_M}}_{U_{M,2}(V_T)} \cdot \mathbbm{1}_{[(1 + m - c )v_0,(1 + m)v_0)}(V_T)  \\
& \,\,\,\,\,+ \underbrace{\frac{1}{1 - b_M}\left( mv_0 + \alpha(V_T - (1 + m)v_0) + a_M \right)^{1-b_M}}_{U_{M,3}(V_T)} \cdot \mathbbm{1}_{[(1 + m)v_0,+\infty)}(V_T).
\end{split}
\end{equation}

We apply Lemma \ref{lem:concavification_3_pieces} to construct the concave envelope of $U_M$ and Theorem \ref{th:optim_w_linear_part_1} to derive the optimal terminal value of the hedge fund.

\begin{corollary}[Fund's optimal terminal value]\label{cor:fl3_optimal_terminal_vaule} $ $\\
Let $U_M$ be as defined in (\ref{eq:utility_hara_manager_fl3}). Denote:
\begin{equation}\label{eq:cases_determinant_ratio}
H = \frac{(mv_0 + a_M)^{1 - b_M} - (v_0(m - c) + a_M)^{1 - b_M}}{(1 - b_M)(1 + m)v_0}
\end{equation}
and
\begin{equation} \label{eq:parametric_regions}
\begin{aligned}
\mathcal{P}_A = \biggl\{ (m, \alpha, c) \in \mathcal{P}:&\, H < \alpha (mv_0 + a_M)^{-b_M}\biggr\};\\
\mathcal{P}_B  = \biggl\{ (m, \alpha, c) \in \mathcal{P}:& \, \alpha(mv_0 + a_M)^{-b_M} \leq H  \leq (mv_0 + a_M)^{-b_M} \biggr\};\\
\mathcal{P}_C  = \biggl\{ (m, \alpha, c) \in \mathcal{P}:& \,(mv_0 + a_M)^{-b_M} < H \biggr\}.
\end{aligned}
\end{equation}
Then, the fund's optimal terminal value is given by\\
Case A, $(m, \alpha, c) \in \mathcal{P}_A$:
  \begin{equation}\label{eq:case_A_optimal_V}
  \begin{aligned}
    V^*_T =& \left( \alpha^{1/b_M - 1}\left(y^*\tilde{Z}_T\right)^{- 1/b_M} + (1+m - \alpha^{-1} m)v_0 - \alpha^{-1} a_M \right) \mathbbm{1}_{\left\{ \tilde{Z}_T \in \left(0,\, s(\theta_1^A)/y^*\right]\right\}},
  \end{aligned}
 \end{equation}
     where $\theta_1^A$  is the unique solution of the following equation w.r.t. $v$
       \begin{equation*}
       	\begin{aligned}
 		(\alpha v + (m - \alpha(1+m))v_0 + a_M)^{-b_M}(b_M \alpha v + (m - \alpha(1+m))v_0 + a_M) \\= (v_0(m - c) + a_M)^{1 - b_M}
	\end{aligned}
       \end{equation*}
       on $v \in ((1 + m)v_0, +\infty)$, and $s(\theta_1^A) = \alpha(\alpha \theta_1^A + (m - \alpha(1+m))v_0 +a_M)^{-b_M}$.\\
Case B, $(m, \alpha, c) \in \mathcal{P}_B$:
  \begin{equation}\label{eq:case_B_optimal_V}
  \begin{aligned}
    V^*_T =&\left( \alpha^{1/b_M - 1}\left(y^*\tilde{Z}_T\right)^{- 1/b_M} + (1+m- \alpha^{-1}m)v_0 - \alpha^{-1}  a_M \right)\\
    &\cdot \mathbbm{1}_{\left\{ \tilde{Z}_T \in \left(0,\, \alpha(mv_0 + a_M)^{-b_M}/y^*\right)\right\}} + (1+m)v_0 \\
    & \cdot \mathbbm{1}_{\left\{ \tilde{Z}_T \in \left[\alpha(mv_0 + a_M)^{-b_M}/y^*,\,  s(\theta_1^B)/y^*)\right]\right\}},
  \end{aligned}
 \end{equation}
 where  $\theta_1^B = (1+m)v_0$ and $s(\theta_1^B) = H$.\\
Case C, $(m, \alpha, c) \in \mathcal{P}_C$:
 \begin{equation}\label{eq:case_C_optimal_V}
  \begin{aligned}
    V^*_T =&\left( \alpha^{1/b_M - 1}\left(y^*\tilde{Z}_T\right)^{- 1/b_M} + (1+m - \alpha^{-1} m)v_0 - \alpha^{-1}  a_M \right)\\
    &\cdot \mathbbm{1}_{\left\{ \tilde{Z}_T \in \left(0,\, \alpha(mv_0 + a_M)^{-b_M}/y^*\right)\right\}} \\
    &+ (1+m)v_0 \mathbbm{1}_{\left\{ \tilde{Z}_T \in \left[\alpha(mv_0 + a_M)^{-b_M}/y^*,\,  (mv_0 + a_M)^{-b_M}/y^*)\right]\right\}} \\
    & + \left( \left( y^*\tilde{Z}_T \right)^{-1/b_M} + v_0 - a_M \right) \mathbbm{1}_{\left\{ \tilde{Z}_T \in \left((mv_0 + a_M)^{-b_M}/y^*,\, s(\theta_1^C)/y^* \right]\right\}},
  \end{aligned}
 \end{equation}
 where  $\theta_1^C$  is the unique solution of the following equation w.r.t. $v$
      \begin{equation*}
       (v - v_0 + a_M)^{-b_M}(b_M v - v_0+ a_M) = (v_0(m - c) + a_M)^{1 - b_M}
       \end{equation*}
       on $v \in ((1 + m - c)v_0, (1 + m)v_0)$, and $s(\theta_1^C) = (\theta_1^C - v_0 + a_M)^{-b_M}$.

In all three cases, $y^* > 0$ is the unique solution of the equation $\mathbb{E}\left[ \tilde{Z}_{T}V_T^{*}\right] = v_0$.
\end{corollary}

\begin{remark}
Note that $\theta_1^X,\,X \in \{ A, B, C\}$, is the rightmost concavification point of $U_M(\cdot)$ in the corresponding concavification case, i.e. the rightmost point of the linear part of the concave envelope of $U_M(\cdot)$.
\end{remark}

In the next two propositions we provide the semi-explicit formulas for the parties' value functions. These are needed for computing FBPO fee structures in Problem \eqref{eq:pareto_optimization_problem}.
\begin{proposition}[Manager's value function]\label{prop:value_fct_manager}$ $\\
Let the manager's preferences be determined by $U_M$ as per \eqref{eq:utility_hara_manager_fl3}. Let $y^*$, $\theta_1^X$ and $s(\theta_1^X),X \in \{ A, B, C\},$ be as defined in Corollary \ref{cor:fl3_optimal_terminal_vaule}. Define:
$$\xi_1 = \ExpTwoPar{(1-b_M)b_M^{-1}}{0.5(b_M-1)^2b_M^{-2}}.$$
Then the manager's value function for $V_T^*$ is given by\\
Case A, $(m, \alpha, c) \in \mathcal{P}_A$:
        \begin{flalign*}
           \phi_M(m, \alpha, c) &= U_M(0)\Phi\left( d_2^A(y^*) \right) + \left( 1 - b_M \right)^{-1}(y^*)^{(b_M-1)/b_M}\alpha^{(1-b_M)/b_M}\xi_1 &\\
           & \,\,\,\,\,\, \cdot \PhiDifFin{d_1^A(y^*)}{d_2^A(y^*)}{+(1-b_M^{-1})},&
        \end{flalign*}
\begin{flalign*}
  \text{where } d_1^A(y^*)  = +\infty, d_2^A(y^*)  = \frac{\log \left( y^* /s(\theta_1^A) \right) - \left( r+ 0.5\gamma^2 \right)T}{\gamma\sqrt{T}}; &&
\end{flalign*}
Case B, $(m, \alpha, c) \in \mathcal{P}_B$:
          \begin{flalign*}
           \phi_M(m, \alpha, c) &= U_M(0)\Phi\left( d_3^B(y^*) \right) + \left( 1 - b_M \right)^{-1}(y^*)^{(b_M-1)/b_M}\alpha^{(1-b_M)/b_M}\xi_1 &\\
           & \,\,\,\,\,\,\cdot \PhiDifFin{d_1^B(y^*)}{d_2^B(y^*)}{+(1-b_M^{-1})} &\\
           & \,\,\,\,\,\,+ (1-b_M)^{-1} (mv_0 + a_M)^{1 - b_M} \PhiDifNull{d_2^B(y^*)}{d_3^B(y^*)},&
        \end{flalign*}
  \begin{flalign*}
		\text{ where } d_1^B(y^*) & = +\infty,\quad d_2^B(y^*)  = \frac{\log \left( y^* \alpha^{-1}(mv_0 + a_M)^{b_M}\right) - \left( r+ 0.5\gamma^2 \right)T}{\gamma\sqrt{T}},&\\
       d_3^B(y^*) &= \frac{\log \left( y^* / s(\theta_1^B)\right) - \left( r+0.5\gamma^2 \right)T}{\gamma\sqrt{T}};&
  \end{flalign*}
Case C, $(m, \alpha, c) \in \mathcal{P}_C$:
            \begin{flalign*}
           \phi_M(m, \alpha, c) &= U_M(0)\Phi\left( d_4^C(y^*) \right) + \left( 1 - b_M \right)^{-1}(y^*)^{(b_M-1)/b_M}\alpha^{(1-b_M)/b_M}\xi_1 &\\
           & \,\,\,\,\,\,\cdot \PhiDifFin{d_1^C(y^*)}{d_2^C(y^*)}{+(1-b_M^{-1})} &\\
           & \,\,\,\,\,\,+ (1-b_M)^{-1} (mv_0 + a_M)^{1 - b_M} \PhiDifNull{d_2^C(y^*)}{d_3^C(y^*)} &\\
           & \,\,\,\,\,\, + (1-b_M)^{-1}  (y^*)^{(b_M-1)/b_M}\xi_1&\\
           &\,\,\,\,\,\, \cdot \PhiDifFin{d_3^C(y^*)}{d_4^C(y^*)}{+(1-b_M^{-1})},
        \end{flalign*}
   \begin{flalign*}
\text{ where } d_1^C(y^*) &= +\infty,\quad d_2^C(y^*)  = \frac{\log \left( y^*\alpha^{-1}(mv_0 + a_M)^{b_M}\right) - \left( r+ 0.5\gamma^2 \right)T}{\gamma\sqrt{T}},&\\
         d_3^C(y^*) &= \frac{\log \left( y^*(mv_0 + a_M)^{b_M} \right) - \left( r+ 0.5\gamma^2 \right)T}{\gamma\sqrt{T}},\quad d_4^C(y^*)  = \frac{\log \left( y^* / s(\theta_1^C) \right) - \left( r+ 0.5\gamma^2 \right)T}{\gamma\sqrt{T}}.&
  \end{flalign*}
\end{proposition}

Let the investor have a HARA utility function:
\begin{equation*}
 \tilde{U}_I(v) = \frac{1}{1 - b_I} \left( v + a_I\right)^{1 - b_I},\,\, \tilde{U}_I(v): (-a_I, +\infty) \rightarrow \mathbb{R},
\end{equation*}
where $a_I \geq v_0(m - c)$ if $b_I \in (0, 1)$ and $a_I > v_0(m - c)$ if $b_I \in (1, +\infty)$. Such choice of $a_I$ guarantees that the function $U_I(V_T):=\tilde U_I(I(V_T))$ is real-valued for any $V_T \in [0, +\infty)$.

\begin{proposition}[Investor's value function]\label{prop:value_fct_investor}$ $\\
Let $y^*$, $\theta_1^X$ and $s(\theta_1^X), X \in \{ A, B, C\},$ be as defined in Corollary \ref{cor:fl3_optimal_terminal_vaule}. Let the investor's preferences be determined by $U_I$ as defined above.
Then the investor's value function is given in\\
Case A, $(m, \alpha, c) \in \mathcal{P}_A$:
        \begin{flalign*}
           \phi_I(m, \alpha, c)  & = (1 - b_I)^{-1}\left( v_0(c - m) + a_I \right)^{1 - b_I} \Phi(d_2^A(y^*)) \\
           & \,\,\,\,\,\, + (1 - b_I)^{-1}\E{\left(k\tilde{Z}_T^{-1/b_M} + l\right)^{1 - b_I}\indV{\tilde{Z}_T\in \left(0, s(\theta_1)/y^*\right]}};&
        \end{flalign*}
Case B, $(m, \alpha, c) \in \mathcal{P}_B$:
          \begin{flalign*}
              \phi_I  (m, \alpha, c) & = (1 - b_I)^{-1}\left( v_0(c - m) + a_I \right)^{1 - b_I} \Phi(d_3^B(y^*)) &\\
              & \,\,\,\,\,\, + (1 - b_I)^{-1}\E{\left(k\tilde{Z}_T^{-1/b_M} + l\right)^{1 - b_I}\indV{\tilde{Z}_T\in (0, \alpha(mv_0 + a_M)^{-b_M}/y^*)}}&\\
 & \,\,\,\,\,\, + (1 - b_I)^{-1}\left( v_0 + a_I \right)^{1 - b_I} \PhiDifNull{d_2^B(y^*)}{d_3^B(y^*)};&
           \end{flalign*}
Case C, $(m, \alpha, c) \in \mathcal{P}_C$:
            \begin{flalign*}
                 \phi_I(m, \alpha, c) &= (1 - b_I)^{-1}( v_0(c - m)+ a_I )^{1 - b_I} \Phi(d_4^C(y^*)) &\\
 & \,\,\,\,\,\, + (1 - b_I)^{-1}\E{\left(k\tilde{Z}_T^{-1/b_M} + l\right)^{1 - b_I}\indV{\tilde{Z}_T\in (0, \alpha(mv_0 + a_M)^{-b_M}/y^*)}} &\\
 & \,\,\,\,\,\, + (1 - b_I)^{-1} \left( v_0 + a_I \right)^{1 - b_I} \PhiDifNull{d_2^C(y^*)}{d_4^C(y^*)},
            \end{flalign*}
where $k = (1 - \alpha)\alpha^{1/b_M - 1}(y^*)^{-1/b_M}$, $l = (1+m - \alpha^{-1}m)v_0  + a_M(1 - \alpha^{-1}) + a_I$, and $d_2^A(\cdot),\, d_2^B(\cdot),\, d_3^B(\cdot), \, d_2^C(\cdot),\, d_4^C(\cdot)$ are defined in Proposition \ref{prop:value_fct_manager}.
\end{proposition}

In \ref{app:aux_proofs}, we provide several supplementary results related to the fund's optimal terminal value. In Proposition \ref{prop:equations_for_y} we derive the explicit form of the equations for finding $y^*$. In Proposition \ref{prop:expectation_of_V_T} we provide the explicit form of the the first and the second moment of $V_T^*$. These are needed for computing the hedge-fund's Sharpe ratio in Problem \eqref{eq:eq_fee_opt_problem}.

\section{Numerical Analysis}\label{sec:hara_numerical_studies}
\subsection{Algorithm Overview}\label{subsec:algorithm}
The selection process for the preferred fee structure is an optimization problem that has several nested optimization subproblems:
\begin{equation}\label{eq:PF_selection_summary}\tag{$P_{SR}$}
  \begin{aligned}
    \max_{m, \alpha, c} &\quad SR(V_T^*(m, \alpha, c)); \\
    \text{s.t.} &\quad V_T^*(m,\alpha, c) \text{ solves } \eqref{eq:nonc_optim_wealth_problem_man} \\
    &\quad \exists \phi_{min} \text{ s.t. }(m,\alpha, c) \text{ solves } \eqref{eq:pareto_optimization_problem}. \\
  \end{aligned}
\end{equation}
To solve some subproblems we resort to numerical techniques, as the analytic solution is not available. Here we provide the algorithm we use in the next subsections to compute the solution to Problem \eqref{eq:PF_selection_summary}:
\begin{enumerate}
	\item Initialize:
	\begin{enumerate}
		\item model parameters $r, \gamma, T, v_0, a_M, b_M, a_I, b_I$;
		\item discretization steps $\Delta m$, $\Delta \alpha$, $\Delta c$, $\Delta \phi_{min}$;
	    \item set $\mathcal{P}_{FBPO} = \emptyset$.
	\end{enumerate}
	\item For each $m \in \{0\%, \Delta m, 2\Delta m, \dots , 5\%\} =: G_m$, $\alpha \in \{\Delta \alpha, 2\Delta \alpha, 3\Delta \alpha, \dots , 50\%\}=:G_{\alpha}$,\\$c \in \{0\%, \Delta c, 2\Delta c, \dots , 30\%\}=:G_c$ calculate:
	\begin{enumerate}
		\item $y^*$ using Proposition \ref{prop:equations_for_y} from \ref{app:aux_proofs} and the bisection method;
		\item $\phi_M(m, \alpha, c)$ using  Proposition \ref{prop:value_fct_manager};
		\item $\phi_I(m, \alpha, c)$ using Proposition \ref{prop:value_fct_investor}.
	\end{enumerate}
		\item Calculate $\displaystyle \phi_{M, min} = \min_{m \in G_m, \alpha \in G_{\alpha}, c \in G_c}\{\phi_M(m, \alpha, c)\}$, $\displaystyle \phi_{M, max} = \max_{m \in G_m, \alpha \in G_{\alpha}, c \in G_c}\{\phi_M(m, \alpha, c)\}$.
	\item For each $\phi_{min} \in \left\{\phi_{M, min} , \phi_{M, min}  + \Delta \phi_{min}, \phi_{M, min}  + 2\Delta \phi_{min}, \dots, \phi_{M, max}   \right\}$ 
	\begin{enumerate}
		\item find a good initial fee structure $(m_{0}, \alpha_{0}, c_{0})$ by solving \eqref{eq:pareto_optimization_problem} on the discrete set  $G_m \times G_{\alpha} \times G_c$;
		\item calculate $(m_{\phi_{\min}}^*, \alpha_{\phi_{\min}}^*, c_{\phi_{\min}}^*)$ that solves \eqref{eq:pareto_optimization_problem} with SQP starting from $(m_{0}, \alpha_{0}, c_{0})$;
	    \item $\mathcal{P}_{FBPO} = \mathcal{P}_{FBPO} \cup (m_{\phi_{\min}}^*, \alpha_{\phi_{\min}}^*, c_{\phi_{\min}}^*)$.
	\end{enumerate}
	\item Find $(\hat m, \hat \alpha, \hat c) \in \mathcal{P}_{FBPO}$ that solves  \eqref{eq:eq_fee_opt_problem} using Proposition \ref{prop:expectation_of_V_T} from \ref{app:aux_proofs}.
\end{enumerate}
Note: Step 4 (a) speeds up the SQP algorithm and decreases the odds of getting to a local optimum.

\subsection{Model Parametrization}
The highest management fee we are aware of equals $5\%$ and was charged at Renaissance Technologies, according to an article published on March 7, 2016, on Investopedia\footnote{https://www.investopedia.com/articles/investing/030716/jim-simons-justifying-5-management-fee.asp}. According to the same source, this hedge fund charged a $44\%$ incentive fee. We find it very improbable that an investor would be willing to pay a management fee of more than $5\%$ of her initial endowment or a performance fee greater than $50\%$ of her net profit. Therefore, we set the upper bound for $m$ and $\alpha$ to $5\%$ and $50\%$ respectively. \cite{Djerroud2016} examined first-loss coverage guarantees between $1\%$ and $25\%$.
To be a bit more flexible, we consider $c \in [0\%, 30\%]$.

We choose the same values of the financial market parameters as those considered in two papers related to the first-loss scheme.  \cite{He2018} conduct numerical studies for an interest rate of $5\%$, arguing that such choice is motivated by historical data. However, they also investigate the case $r = 2\%$ due to the low interest-rate environment and conclude that the results for the low interest rate are similar to those obtained for $r = 5\%$. \cite{Djerroud2016} consider $r = 2\%$. To be in line with the mentioned papers and consistent with the current economic conditions in Europe and North America, we set $r = 2\%$. We fix the market price of risk $\gamma$ at $40\%$, as it is done in \cite{He2018}. Since the majority of hedge funds charge fees annually, we set $T = 1$. We assume that $v_0 = 1$.

For many years the hedge-fund sector was asking for management fees around $2\%$ and performance fees around $20\%$. However, due to the investors' concerns that such fees might not be justified by the hedge-funds' performance, the management fees have been decreasing. In May 2018, the average management fee was traded around $1.58\%$\footnote{http://docs.preqin.com/newsletters/hf/Preqin-Hedge-Fund-Spotlight-May-2018.pdf}, whereas some hedge funds had already canceled their management fees completely. We believe that the hedge-fund sector will increase in transparency and charged fees will approach the optimal ones. In our view, hedge funds that stick to the traditional fee structure will thus gradually end up with a management fee close to $0\%$ and a performance fee around $20\%$. Therefore, we assume that these values of the traditional fee structure parameters are optimal in the absence of the first-loss coverage guarantee and best represent risk preferences of an average investor and an average manager of a hedge fund. Taking into account the technical conditions
\begin{equation*}
  \begin{aligned}
      a_I \stackrel{(\ref{eq:terminal_wealth_investor})}{\geq}  \max_{\scriptsize \begin{matrix} m \in [0, 0.05] \\  c \in [0, 0.3] \end{matrix}}\{v_0(m - c)\} = 0.05,\qquad  a_M \stackrel{(\ref{eq:terminal_wealth_manager})}{\geq}  \max_{\scriptsize \begin{matrix} m \in [0, 0.05] \\  c \in [0, 0.3] \end{matrix}}\{v_0(c - m)\} = 0.3,
  \end{aligned}
\end{equation*}
which guarantee that the utility function evaluated at the minimal terminal wealth of the corresponding party is finite, we find the following parameters for our HARA utility functions that are consistent with the above mentioned intuition of the  traditional fee structure: $a_M = 0.3,\,b_M = 0.65,\,a_I = 0.3,\, b_I = 0.65$. For these parameters the investor's optimal fee structure without any first-loss coverage guarantee is:
\begin{equation}\label{eq:optimal_trad_fee_investor}
	\argmax_{\scriptsize \begin{matrix} m \in [0\%,\,5\%] \\  \alpha \in [0\%,\,50\%] \end{matrix}} \phi_I(m, \alpha, 0\%) = (0\%, 20.3\%).
\end{equation}

Note that the investor's value function does not appear in the manager's portfolio optimization problem. $\phi_I(m, \alpha, c)$ comes from Proposition \ref{prop:value_fct_investor} and equals the investor's expected utility for the terminal portfolio value optimal for the fund manager who solves (\ref{eq:nonc_optim_wealth_problem_man}) for a certain fee structure $(m, \alpha, c) \in \mathcal{P}$. The obtained risk-aversion parameters are consistent with \cite{Holt2002}. Analyzing power utility functions, which are a subclass of HARA utility functions, the authors classify a decision maker as  (moderately) risk averse if $b_M (b_I) \in (0.41, 0.68)$. We consider $a_M = a_I = 0.3$ and $b_M=b_I=0.65$ as base case parameters in our analysis, i.e. the manager and the investor have the same utility function. Later in the study we will vary these parameters for further investigations.

Parameter specifications in the base case are summarized in Table \ref{tab:base_case_parameters}.

\begin{table}[!h]
\begin{center}
\caption{Values of the model parameters in the base case}\label{tab:base_case_parameters}
  \begin{tabular}{@{} lcccccccc @{}}
    \toprule
    \multicolumn{1}{l}{} &
    \multicolumn{2}{c}{Market} &
    \multicolumn{2}{c}{Investment} &
    \multicolumn{4}{c}{Utility functions} \\
    \cmidrule(lr){2-3} \cmidrule(lr){4-5} \cmidrule(lr){6-9}
    Parameters & $r$ & $\gamma$ & $v_0$ & $T$ & $a_M$ & $b_M$ & $a_I$ & $b_I$ \\
    \midrule
    Value & $2\%$ & $40\%$ & $1$ & $1$ & $0.3$ & $0.65$ & $0.3$ & $0.65$ \\
    \bottomrule
  \end{tabular}
\end{center}
\end{table}

\subsection{From Traditional to First-Loss Fee Structure}\label{subs:traditional_to_fl}
\begin{figure}[!htbp]
\begin{subfigure}{.5\textwidth}
  \centering
  \includegraphics[width=\linewidth]{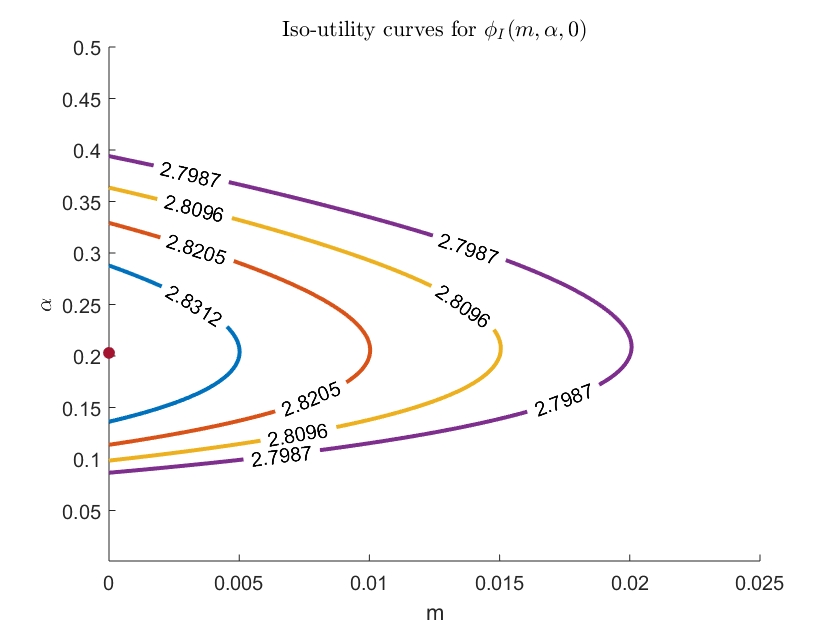}
  \caption{Investor's iso-utility curves in the traditional\\ compensation scheme}
\label{sfig:inv_isocurves_fixed_c}
\end{subfigure}%
\begin{subfigure}{.5\textwidth}
  \centering
  \includegraphics[width=\linewidth]{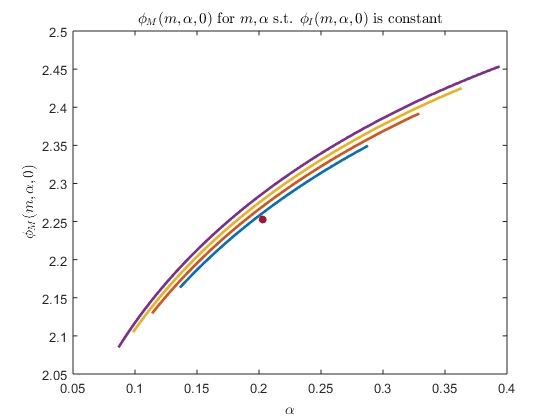}
  \caption{Manager's value function along the investor's iso-utility\\curves given $c = 0$}
  \label{sfig:man_val_fct_along_inv_iso_curves}
\end{subfigure}
\caption{Investor's iso-utility curves in the traditional compensation scheme for iso-utility values $\phi_{I}(2\%,\,20\%,\, 0\%)=2.7987$, $\phi_{I}(1.5\%,\,20\%,\,0\%)=2.8096$, $\phi_{I}(1\%,\,20\%,\, 0\%)\,=\,2.8205$, $\phi_{I}(0.5\%,\,20\%,\, 0\%)\,=\,2.8312$ and the manager's expected utility along them}
\label{fig:inv_isocurves_man_val_fct_fixed_c}
\end{figure}

Let us first consider the traditional fee structure. Setting $c = 0\%$ in our model, we plot in Figure \ref{fig:inv_isocurves_man_val_fct_fixed_c} the investor's  indifference curves corresponding to several traditional fee structures and the manager's expected utility along these iso-utility curves of the investor. The lines in Figure \ref{sfig:inv_isocurves_fixed_c} show all parameter pairs $(m,\alpha)$ that lead to the same expected utility of the investor. The iso-utility values originate from the fees common in the hedge-fund sector: $2\%\,\&\,20\%$ (violet line), $1.5\%\,\&\,20\%$ (yellow line), $1\%\,\&\,20\%$ (red line), $0.5\%\,\&\,20\%$ (blue line). We use the same coloring structure in Figure \ref{sfig:man_val_fct_along_inv_iso_curves}. The trade-off between the management and the performance fees is also analyzed in \cite{Goetzmann2003}. However, the authors work in a valuation framework and consider a hedge fund with two peculiarities -- a liquidation boundary and a traditional compensation structure where the performance fee is based on the hedge-fund's high-water mark.

In Figure \ref{sfig:inv_isocurves_fixed_c}, we observe that the lower the management fee, the more the iso-utility curves approach the constrained maximal value of $\phi_I$, which is marked in brown. 
Note that any change in the fee structure changes $\phi_I$ indirectly: it influences first the manager's optimal portfolio allocation decision, which in turn has an impact on the value function of the investor. Consider for example a manager who charged a $2\% \& 20\%$ fee structure and wants to renegotiate the fee structure while keeping the client as happy as before, i.e. the manager should choose a fee structure on the violet line. For performance fees above $20\%$, we can easily see that the higher the performance fee, the lower the management fee should be to keep the investor's value function constant. For example, an increase of the performance fee from $20\%$ to $30\%$ should be compensated by a decrease of the management fee from $2\%$ to $1.5\%$, so that the investor's expected utility stays at the same level as it has for the $2\%\, \& \, 20\%$ fee structure. According to Figure \ref{sfig:man_val_fct_along_inv_iso_curves}, a rational manager will not negotiate a lower performance fee while keeping the investor's expected utility level constant, as the manager's expected utility decreases in this case.

In contrast to the investor's value function, the manager's value function is strictly increasing in both $m$ and $\alpha$. So if an investor insists on decreasing the management fee, then the manager has to increase the performance fee to maintain her expected utility constant.

Note that the Pareto optimal fees can be also calculated for the traditional fee structure, namely by solving Problem (\ref{eq:pareto_optimization_problem}) for fixed $c=0\%$ and different $\phi_{min}\in \R$. In the numerical studies done for various combinations of $b_M\in (0,1)$ and $b_I \in (0,1)$, we observed that in the universe of traditional fees each Pareto optimal fee has either $m^*= 0\%$ or $\alpha^* = 50\%$. The fee $(2\%,20\%)$ is clearly not Pareto optimal and favors the manager more than the investor. This is compatible with the investors' concerns regarding high fees that may not be justified by the hedge-funds' performance and the push for a decrease in management fees given a fixed $20\%$ performance fee.

As already mentioned, in the traditional fee structure the manager always earns at least the management fee, whereas the investor is not provided with any guarantee regarding her minimal profit or maximal loss. This asymmetry, along with other reasons such as poor performance of hedge funds in comparison to passive investment vehicles, animated some hedge funds to start offering first-loss coverage guarantees along with management and performance fees. But how much are investors better-off under the new fee structure? As market management fees are \enquote{converging} to $0\%$, what is the trade-off between performance fees and the first-loss coverage guarantee levels?

\begin{figure}[!ht]
\begin{center}
  \includegraphics[width = 10 cm]{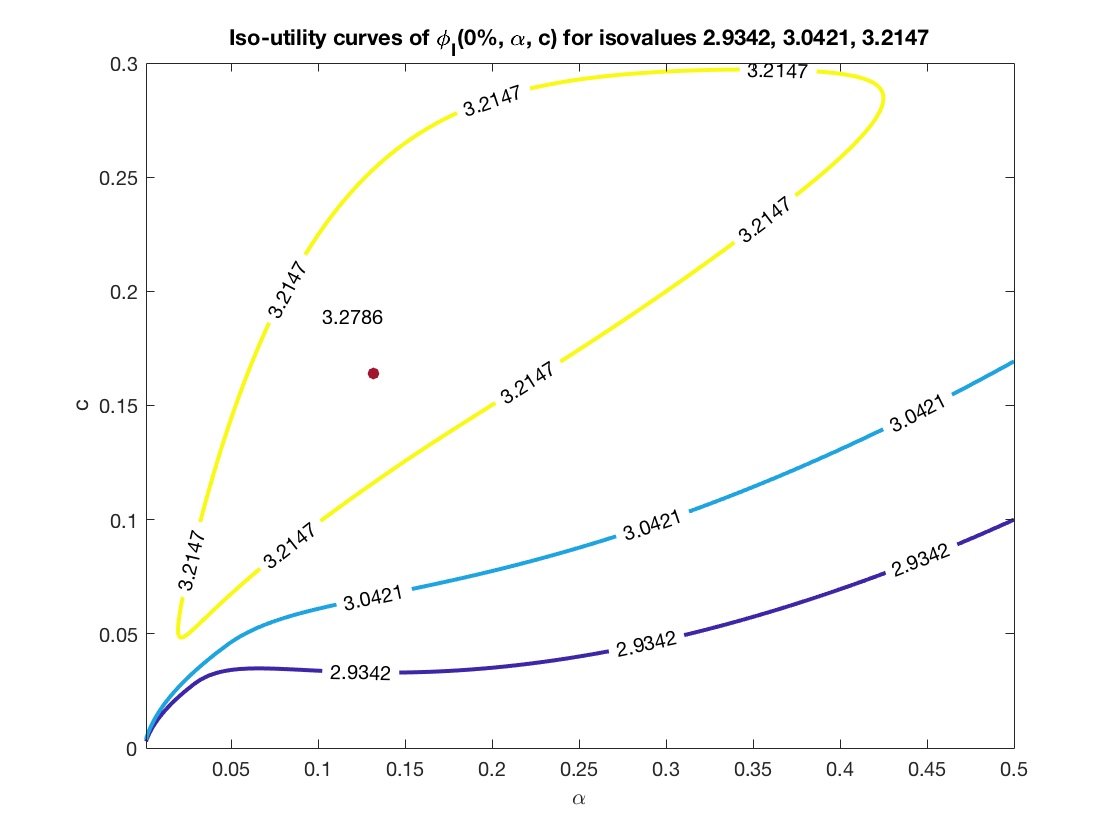}
\caption{Investor's iso-utility curves in the first-loss compensation scheme with $m=0\%$ for iso-utility values $\phi_{I}(0\%,\,50\%,\,10\%)=2.9342$, $\phi_{I}(0\%,\,30\%,\,10\%)=3.0421$, $\phi_{I}(0\%,\,30\%,\,20\%)\,=\,3.2147$}
\label{fig:inv_isocurves_fixed_m}
\end{center}
\end{figure}

To get an intuition to an answer to these questions, we fix $m = 0\%$ and plot several iso-utility curves of the investor in Figure \ref{fig:inv_isocurves_fixed_m}. We see that the trade-off between the performance fee and the first-loss coverage guarantee is intuitive -- the higher performance fee, the higher the first-loss coverage guarantee should be to maintain the investor's expected utility on the same level.
Consider, for example, the violet line in Figure \ref{fig:inv_isocurves_fixed_m} which corresponds to a relatively popular first-loss fee structure $(0\%,\,50\%,\,10\%)$. If the manager decides to decrease the first-loss coverage guarantee from $10\%$ to $5\%$, she should also decrease the performance fee from $50\%$ to $35\%$ in order to maintain the investor's utility level constant. Comparing with Figure \ref{sfig:inv_isocurves_fixed_c}, we also see that the percentage gain in the investor's value function from switching to this fee from the optimal traditional fee $(0\%,20\%,0\%)$ is about $3\%$. The blue line corresponds to the performance fee $30\%$ and the first-loss coverage guarantee $10\%$, which are recommended for the first-loss scheme in \cite{He2018}. Under this first-loss fee structure, the investor is further better off, although she is still far from her maximal expected utility. Further increase in the first-loss coverage guarantee to $20\%$, while other fee parameters are fixed, results in a significant increase in the investor's value function.
The maximum of the investor's value function (marked in brown) is attained at $\alpha = 13.2\%,\,c=16.4\%$ and equals $3.2786$, which is $15\%$ larger than the expected utility for the optimal fee in the traditional setting.

 \begin{figure}[!ht]
\begin{subfigure}{.5\textwidth}
  \centering
  \includegraphics[width=.95\linewidth]{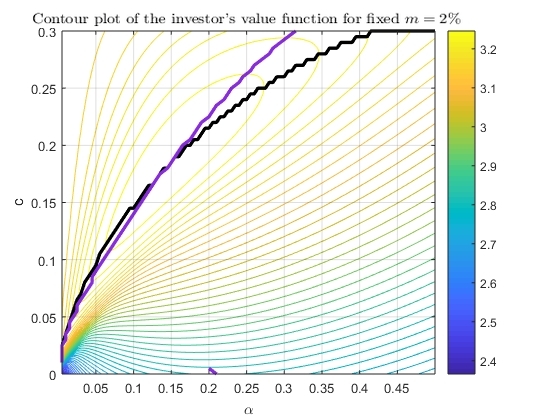}
  \caption{ $ m = 2\% $ }
  \label{sfig:inv_heatplot_fix_m_2}
\end{subfigure}%
\begin{subfigure}{.5\textwidth}
  \centering
  \includegraphics[width=.95\linewidth]{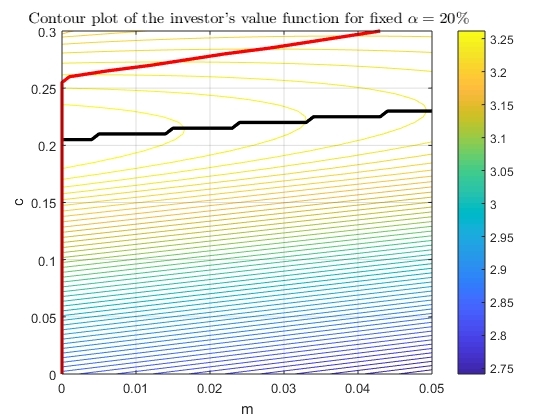}
  \caption{ $ \alpha = 20\% $ }
  \label{sfig:inv_heatplot_fix_alpha_20}
\end{subfigure}
\vskip \baselineskip
\begin{subfigure}{.5\textwidth}
  \centering
  \includegraphics[width=.95\linewidth]{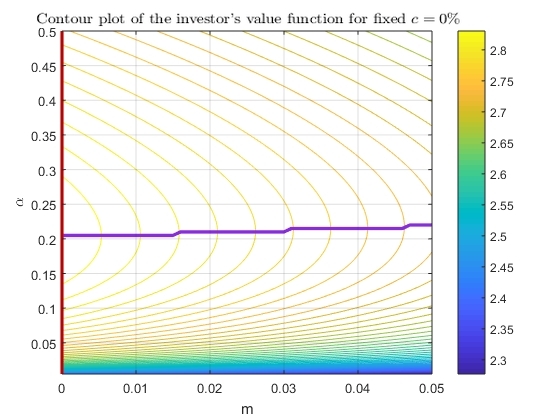}
  \caption{ $ c = 0\% $ }
  \label{sfig:inv_heatplot_fix_c_0}
\end{subfigure}%
\begin{subfigure}{.5\textwidth}
  \centering
  \includegraphics[width=.95\linewidth]{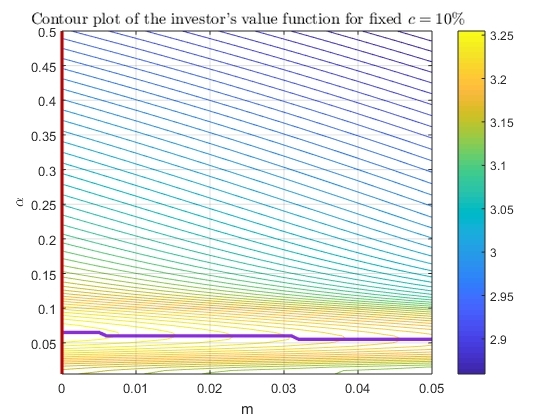}
  \caption{ $ c = 10\% $ }
  \label{sfig:inv_heatplot_fix_c_10}
\end{subfigure}
\caption{Investor's value function in the base case.} 
\label{fig:inv_heatplot_fix_params}
\end{figure}

Figure \ref{fig:inv_heatplot_fix_params} illustrates the contour plots of the investor's value function. Each subfigure contains 50 equidistant contours (thin lines). In each subfigure, two thick lines indicate the constrained maximum of the investor's value function w.r.t one of the variables as a function of another variable given that the value of the third variable is fixed as indicated in the legend. These lines are related to the constrained optimal $m$ (red), $\alpha$ (purple) and $c$ (black) respectively.
We see in Subfigure \ref{sfig:inv_heatplot_fix_m_2} that the risk-averse investor is willing to pay a performance fee higher than the minimal one. This differs from the findings in \cite{He2018}, who discover that the expected utility of a loss-averse investor is decreasing in the performance fee.  If $\alpha$ is low, the manager is not incentivized well enough to generate attractive returns for the hedge fund. Consequently, the investor is worse off. Obviously, high performance fees have also a negative impact on the investor's terminal wealth. All that leads to a moderate optimal $\alpha$ strictly positive and larger than its minimal value.

We also observe in Subfigure \ref{sfig:inv_heatplot_fix_m_2} and  in Subfigure \ref{sfig:inv_heatplot_fix_alpha_20} that the investor would negotiate a moderate first-loss coverage guarantee. Low levels of the first-loss coverage guarantee (or its absence) are good for the manager. For a fixed management fee, the first-loss coverage guarantee is an increasing function of the performance fee and vice versa. See Subfigure \ref{sfig:inv_heatplot_fix_m_2}. However, high levels of the first-loss coverage guarantee motivate the manager to take less risks, as she is responsible for potential losses with her own money. Lower risks are accompanied by lower returns, which, in turn, decrease the investor's terminal wealth as well.

\begin{figure}[!ht]
\begin{subfigure}{.5\textwidth}
  \centering
  \includegraphics[width=\linewidth]{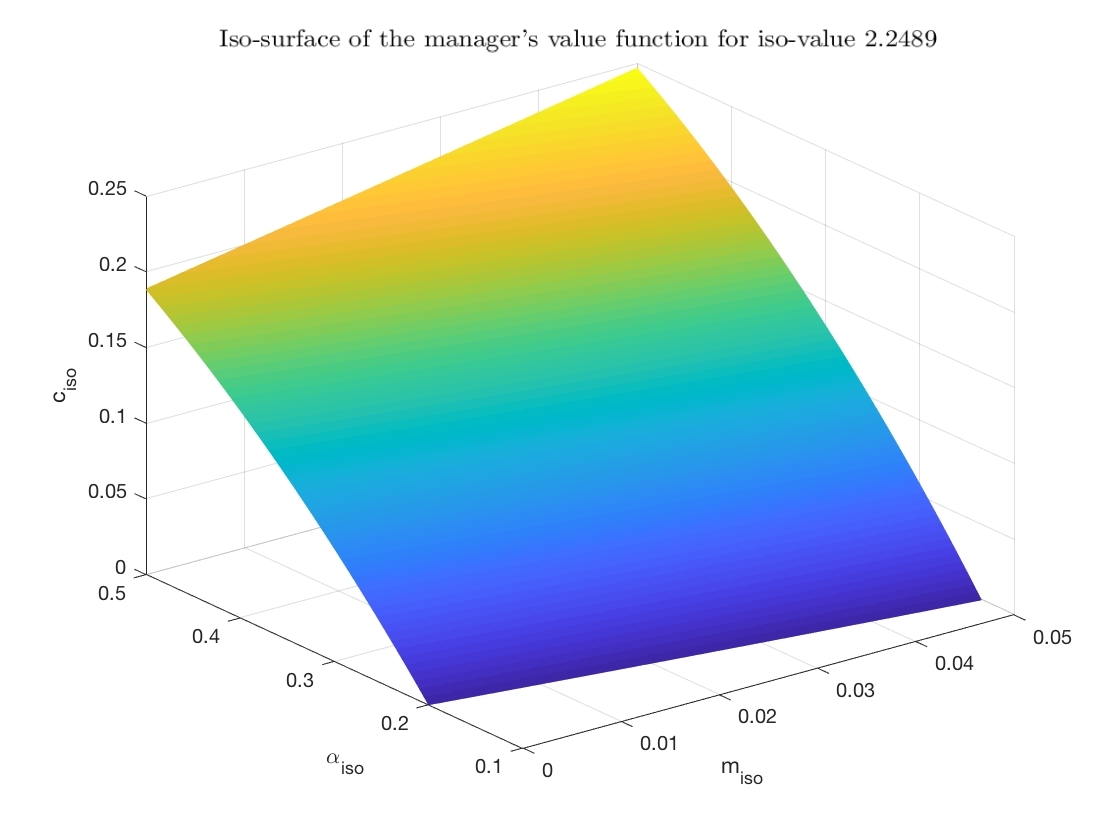}
  \caption{Manager's iso-utility surface for isovalue \\\hspace{\textwidth} $\phi_M(0\%, 20\%, 0\%)=2.2489$}
  \label{sfig:iso_surface_phi_M_0_20_0}
\end{subfigure}%
\begin{subfigure}{.5\textwidth}
  \centering
  \includegraphics[width=\linewidth]{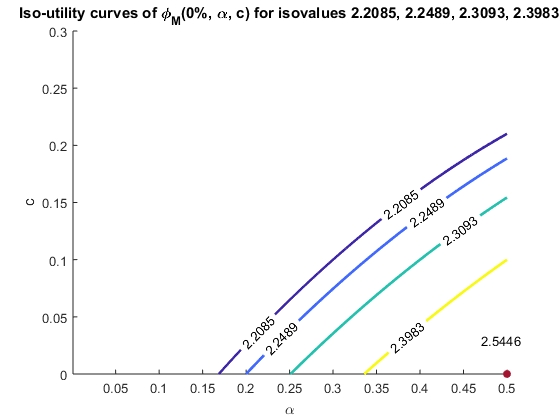}
  \caption{ Iso-utility curves of $\phi_M(0\%, \alpha, c)$ for various isovalues}
  \label{sfig:man_isocurves_m_2}
\end{subfigure}
\caption{Indifference regions of the manager.}
\label{fig:man_iso}
\end{figure}

On the contrary to the investor's expected utility, the manager's value function is strictly increasing in $m$ and $\alpha$, and strictly decreasing in $c$. So the manager's highest expected utility is attained at $m = 5\%,\,\alpha = 50\%, \, c = 0\%$. To see the trade-off between fee constituents from the manager's perspective, we illustrate her indifference regions in Figure \ref{fig:man_iso}. In Subfigure \ref{sfig:iso_surface_phi_M_0_20_0} we plot the set of all first-loss fee structures for which the manager has the same expected utility as she does for the fee structure $(0\%, 20\%, 0\%)$. Consider a situation when the investor insists on having a $20\%$ first-loss coverage guarantee. Then the manager has to increase $m$ to $1\%$ and $\alpha$ to $50\%$ to get the same expected utility as she has for the traditional fee structure. Subfigure \ref{sfig:man_isocurves_m_2} shows the iso-utility curves of the manager for fixed $m=0\%$. The isovalues originate from fee structures mentioned in  \cite{Djerroud2016} and \cite{He2018}: $\phi_M(0\%, 30\%, 10\%) = 2.2085,\,\phi_M(0\%, 20\%, 0\%) = 2.2489, \phi_M(0\%, 40\%, 10\%) = 2.3093, \phi_M(0\%, 50\%, 10\%)=2.3983$. We see that the traditional fee structure $(0\%, 20\%, 0\%)$ yields a higher expected utility to the manager (in the base case) than the fee arrangement $(0\%, 30\%, 10\%)$\footnote{ \cite{He2018} find that this fee structure often yields higher expected utilities to both loss averse managers and loss averse investors in a hedge fund with parties money being commingled}. However,  it yields her a lower expected utility than the first-loss fee structures $(0\%, 40\%, 10\%)$ and $(0\%, 50\%, 10\%)$.
The manager's expected utility can be improved further by charging maximal performance fee and offering minimal first-loss coverage guarantee (see brown mark). For the manager, the lower the performance fee she charges, the lower first-loss coverage guarantee she should offer to keep her utility level constant. For example, assume that the investor being charged by the fee structure $(0\%, 50\%, 10\%)$ (yellow line) insists on a lower performance fee, for example $40\%$.  The manager, to preserve her expected utility, should decrease the  first-loss coverage guarantee from $10\%$ to $4\%$.

\subsection{First Best Pareto Optimal Fee Structures and Fee Preferences}\label{subsec:pareto_optimality}
In Subsection \ref{subs:traditional_to_fl} we have seen that each party has different preferences regarding fee structures. Next we calculate FBPO fee structures and find the single optimal fee structure that maximizes the hedge-fund's Sharpe ratio.  For each admissible $\phi_{min}$, we calculate the corresponding FBPO fee structure using the SQP approach with a good starting point (see Step 4 in the algorithm in Subsection \ref{subsec:algorithm}).

\begin{figure}[!htbp]
\begin{subfigure}{.5\textwidth}
  \centering
  \includegraphics[width=\linewidth]{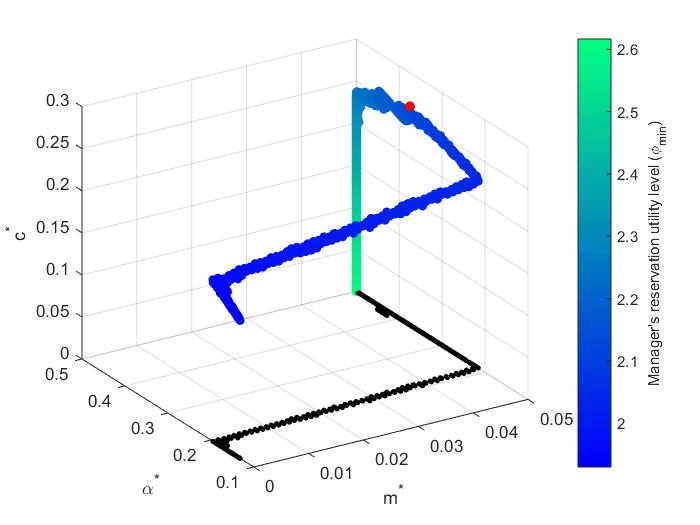}
  \caption{Set of first best Pareto optimal fee structures}
  \label{sfig:pareto_efficient_fees}
\end{subfigure}%
\begin{subfigure}{.5\textwidth}
  \centering
  \includegraphics[width=\linewidth]{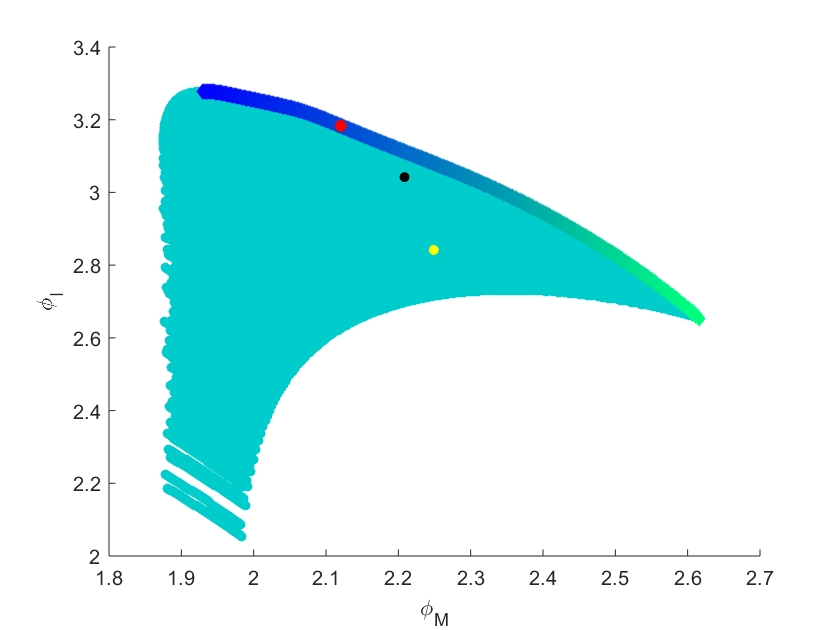}
  \caption{Attainable expected utilities and Pareto frontier}
  \label{sfig:pareto_frontier}
\end{subfigure}
\caption{First best Pareto optimal fee structures and Pareto frontier in the base case}
\label{fig:pareto_efficient_fees}
\end{figure}

Figure \ref{fig:pareto_efficient_fees} illustrates the set of FBPO fee structures, the parties' attainable expected utilities and the Pareto frontier in the base case. First, consider Subfigure \ref{sfig:pareto_efficient_fees} and in particular the plot of the parametric curve $\mathcal{C}:=\mathcal{C}(\phi_{\min})=(m_{\phi_{\min}}^*, \alpha_{\phi_{\min}}^*, c_{\phi_{\min}}^*)$ showing FBPO fee structures depending on $\phi_{\min}$. The subfigure indicates that $\mathcal{C}(\phi_{\min})$ is continuous w.r.t. $\phi_{\min}$ and has four regions where FBPO fee structures behave differently w.r.t. changes in the manager's reservation utility level.
In the front end of the plot of $\mathcal{C}$ we see the FBPO fee structure  $(0\%,14\%, 17\%)$ that corresponds to the manager's minimal reservation utility level $\phi_{\min} = 1.925$.  We observe that for small $\phi_{\min}$, FBPO fee structures have zero management fee but non-trivial performance fee and first-loss coverage guarantee. As the manager's reservation utility level increases, it is not possible to satisfy her participation constraint with zero management fee. Consequently, FBPO fee structures for moderate values of $\phi_{min}$ are non-trivial in all components. For high reservation utility levels of the manager, the optimal management fee attains its maximal value of $5\%$. The optimal performance fee is growing and reaches its maximum as well, whereas the first-loss coverage guarantee is still non-trivial. The largest $\phi_{\min}$ leads to the FBPO fee structure $(5\%,50\%,0\%)$, depicted in the back end of the plot of $\mathcal{C}$. 

Let us now analyze the behavior of each component of FBPO fee structures separately. Consider the projection (black dots) of the FBPO fee structures on the $(m^*,\alpha^*)$--plane. Looking at it from the front left to the rear right,  we observe that for increasing $\phi_{\min}$ the optimal management fee is non-decreasing. In a similar way we verify that the optimal performance fee is also non-decreasing w.r.t. $\phi_{min}$. On the contrary, looking at the whole plot of $\mathcal{C}$, we find that the optimal first-loss coverage guarantee as a function of $\phi_{min}$ is not monotonic. It is increasing up to a certain level that is strictly smaller than the maximal possible one ($c_{\phi_{min}}^* = 26.2\%$ for moderate participation constraint of the manager $\phi_{min}\approx 2.2$). After that, $c_{\phi_{min}}^*$ decreases with increasing $\phi_{min}$. When $\phi_{min}$ is large, the highest management and performance fees along with some positive first-loss guarantee are not satisfactory for the manager. Hence, the only way the investor could appease the manager's appetite in such cases would be decreasing the first-loss coverage guarantee.

We also observe that in the first-loss setting common fees used in the traditional fee structure  (i.e. $m \in [0\%,\,2\%],\,\alpha \in [15\%, 25\%],\, c= 0\%$) are not Pareto optimal. The majority of FBPO fee structures have $c^* \in [15\%, 25\%]$ and they correspond to the situations when the manager's participation condition is moderate, i.e. it is neither too restrictive nor too slack. The point marked red is the the preferred fee structure. It has a management fee of $5\%$, a performance fee of $37.5\%$ and a first-loss coverage guarantee of $26\%$. Interestingly, this first-loss coverage guarantee is slightly higher than the first-loss coverage guarantee levels commonly used in the hedge-fund industry ($10\%-20\%$).

Consider now Subfigure \ref{sfig:pareto_frontier} showing the parties' attainable expected utilities and the Pareto frontier. The frontier is colored in accordance with Subfigure \ref{sfig:pareto_efficient_fees}. The point marked yellow originates from the fee structure $(0\%, 20\%, 0\%)$, whereas the point colored in black originates from the fee structure $(0\%,  30\%, 10\%)$, recommended in \cite{He2018}. We observe that both fee structures are not Pareto optimal in our hedge-fund model, although the latter scheme yields the parties' expected utilities much closer to the Pareto frontier. Note that our hedge-fund model differs from the one considered in \cite{He2018} (manager's and the investor's money commingled, liquidation boundary), which might explain why the black point is not on the efficient frontier. As before, the red marker corresponds to the preferred fee structure, which the manager and the investor should eventually agree on in the framework we consider.

\subsection{Sensitivity Analysis of Preferred Fee Structures}\label{sec:sensitivity_optimal}
In this subsection we explore the impact of various model parameters on the preferred fee structures. First, we investigate the influence of the risk-aversion parameters  $b_M$ and $b_I$ on $(\hat{m}, \hat{\alpha}, \hat{c})$. The preferred fee structures for $b_M$ and $b_I$ taking values in the set $\{0.35,\, 0.45,\, 0.55,\, 0.65,\,0.75,\,1.25,\,2.5,\,5\}$ are shown in Table \ref{tab:unc_pref_fee_structures_bM_bI}. The calculated preferred fee structures show clear patterns.

\begin{sidewaystable} 
\centering
\caption{Preferred fee structures $(\hat m, \hat \alpha, \hat c)$  (in percent) for various coefficients of risk aversion}\label{tab:unc_pref_fee_structures_bM_bI} 
\begin{tabular}{lccccccccc}\toprule
& $b_I = 0.35$ & $b_I = 0.45$& $b_I = 0.55$& $b_I = 0.65$& $b_I = 0.75$ & $b_I = 1.25$ & $b_I = 2.5$ & $b_I = 5$  \\\midrule
$b_M = 0.35$&  $(4.9, 33.7, 26.2)$ & $(4.9, 39.3, 30.0)$ & $(4.1, 40.8, 30.0)$ & $(1.9, 44.7, 30.0)$ & $(0.3, 45.7 , 30.0)$ & $(0.0, 47.0, 30.0)$ & $(0.0, 47.0, 30.0)$ & $(0.0, 47.0, 30.0)$ \\
$b_M = 0.45$ & $(5.0, 27.9, 21.4)$ & $(5.0, 34.7, 26.4)$ & $(5.0, 39.1, 29.6 )$ & $(4.4, 46.5, 30.0)$ & $(4.0, 46.0, 30.0)$ & $(0.1, 50.0, 30.0)$ & $(0.1, 50.0, 30.0)$ & $(0.1, 50.0, 30.0)$ \\
$b_M = 0.55$  & $(5.0, 26.9, 17.9)$ & $(5.0, 29.0, 23.0)$ & $(5.0, 36.0, 26.3)$ & $(5.0, 40.5, 29.0)$ & $(5.0, 43.0, 30.0)$ & $(1.6, 49.8, 30.0)$ & $(0.1, 50.0, 30.0)$ & $(0.1, 50.0, 30.0)$ \\
$b_M = 0.65$ &  $(5.0,  22.0, 15.5)$ & $(5.0, 27.6, 20.4)$ & $(5.0, 33.1, 23.8)$ & $(5.0, 37.5,  26.0)$ & $(5.0, 40.1 , 28.2)$ & $(4.5, 46.0,  30.0)$ & $(1.2, 50.0, 30.0)$ & $(1.2, 50.0, 30.0)$  \\
$b_M = 0.75$ & $(5.0, 19.5, 13.5)$ & $(5.0, 24.7, 18.3)$ & $(5.0, 30.0, 21.4)$ & $(5.0, 34.5, 24.0)$ & $(5.0, 38.8, 26.4)$ & $(4.8,  47.5, 30.0)$ & $(2.6, 50.0, 30.0)$ & $(2.6, 50.0, 30.0)$ \\
$b_M = 1.25$ & $(5.0, 9.0, 8.5)$ & $(5.0, 14.0, 11.5)$ & $(5.0, 19, 14.3)$ & $(5, 23.8, 17.0)$ & $(5.0, 29.6, 19.4)$ & $(5.0, 42.5, 25.0)$ & $(5.0, 50.0, 26.5)$ & $(5.0, 50.0, 26.5)$ \\
$b_M = 2.5$ & $(5.0, 4.5, 4.5)$ & $(5.0, 7.4, 5.8)$ & $(5.0, 10.0, 7.7)$ & $(5.0, 12.5, 9.5)$ & $(5.0, 14.0, 10.5)$ & $(5.0, 23.0, 15.0)$ & $(4.7, 50.0, 30.0)$ & $(4.7, 50.0, 30.0)$ \\
$b_M = 5$  & $(5.0, 2.5, 2.5)$ & $(5.0, 3.12, 2.7)$ & $(5.0,  4.8, 3.8)$ & $(5.0, 5.5, 4.5)$ & $(5.0, 6.6, 5.3)$ & $(5.0, 12.6, 8.9)$ & $(4.7, 50.0, 27.0)$ & $(4.7, 50.0, 27.5)$ \\
\bottomrule
\end{tabular}
\bigskip
\bigskip
\caption{Preferred fee structures $(\hat{m}, \hat{\alpha}, \hat{c})$ (in percent) for various interest rates ($r$)}\label{tab:unc_pref_fee_structures_r}
\begin{tabular}{lccccc}
\toprule
 & $r = -2\%$ & $r = 0\%$ & $r = 2\%$ & $r = 4\%$ & $r = 6\%$ \\ 
\midrule 
$b_M = b_I = 0.65$ & $(5.00, 38.73, 25.95)$ & $(5.00, 35.50, 26.00)$ & $(5.00, 37.49, 26.01)$ & $(5.00, 37.04, 26.21)$ & $(5.00, 37.49, 26.54)$ \\ 
$b_M = b_I = 2.5$ & $(2.91, 50.00, 30.00)$ & $(3.81, 50.00, 30.00)$ & $(4.70,  50.00, 30.00)$ & $(5.00, 50.00, 19.01)$ & $(5.00, 50.00, 18.51)$ \\ 
\bottomrule 
\end{tabular} 
\bigskip
\bigskip
\caption{Preferred fee structures $(\hat{m}, \hat{\alpha}, \hat{c})$ (in percent) for various market prices of risk ($\gamma$)}\label{tab:unc_pref_fee_structures_gamma}
\begin{tabular}{lcccccc}
\toprule
 & $\gamma = 30\%$ & $\gamma = 40\%$ & $\gamma = 50\%$ & $\gamma = 60\%$ & $\gamma = 70\%$ \\ 
\midrule 
$b_M = b_I = 0.65$  & $(5.00, 34.00, 26.00)$ & $(5.00, 37.49, 26.01)$ & $(5.00, 40.92, 25.08)$ & $(5.00, 44.41, 23.78)$ & $(5.00, 44.04, 22.52)$ \\ 
$b_M = b_I = 2.5$  & $(4.99, 50.00, 30.00)$ & $(4.70,  50.00, 30.00)$ & $(5.00, 50.00, 20.01)$ & $(5.00, 50.00, 19.51)$ & $(5.00, 50.00, 20.01)$ \\ 
\bottomrule 
\end{tabular}
\end{sidewaystable} 

For a fixed value of the manager's risk aversion parameter $b_M$ and increasing investor's risk aversion parameter $b_I$, we observe that $\hat m$  is decreasing, $\hat \alpha$ is increasing, $\hat c$ is increasing.
Hence, a more risk-averse investor prefers a lower management fee, a higher performance fee, and a larger first-loss coverage guarantee.

For a fixed value of the investor's risk aversion parameter $b_I$ and increasing manager's risk aversion parameter $b_M$, we observe that $\hat m$ has an increasing,  $\hat \alpha$ has a decreasing,  $ \hat c$ has a decreasing trend. So the more risk-averse the manager is, the higher the management fee, the lower the performance fee and the lower first-loss coverage guarantee tend to be. 

The majority of obtained $\hat m$ are higher than the management fees usually seen in the industry. The majority of obtained $\hat \alpha$ are close to actually traded performance fees of  $20\% - 50\%$ (as part of first-loss fee structures, see e.g. \cite{Djerroud2016}, \cite{He2018}). The computed $\hat c$ are usually higher than industry-common first-loss coverage guarantee levels of  $10\% - 20\%$\footnote{See \url{http://www.ogcap.com/benefits-first-loss-firlo-capital/}}. This difference may be explained by practical peculiarities: transaction costs, taxes, or more complex asset prices dynamics. 

The impact of the interest rate $r$  on the preferred fee structures is shown in Table \ref{tab:unc_pref_fee_structures_r}.  Here we consider decision makers with the same level of risk aversion and consider two cases: $b_M = b_I = 0.65 < 1$ and $b_M = b_I = 2.5 > 1$.  The former level of risk aversion is motivated in Section \ref{sec:hara_numerical_studies}. The latter level of risk aversion $2.5$ is a good compromise between the following two papers. \cite{Koijen2014} find that the median level of risk aversion for mutual funds equals $2.43$ and the mean is equal to $5.72$, \cite{Buraschi2014} conduct numerical studies for risk aversion parameters $1.5$ and $2$. We observe that the preferred management fee tends to increase with the interest rate.
The preferred performance fee does not show any monotonic behavior with respect to $r$ and fluctuates between $35\%$ and $50\%$. The preferred first-loss coverage guarantee exhibits a decreasing trend.
Constant $\gamma$ along with increasing interest rates mean a better risk-return profile of the risky asset $S_1$\footnote{$r \uparrow \Rightarrow \gamma \sigma + r = \mu \uparrow$,\, $r \uparrow \Rightarrow \gamma^{-1}(\mu - r) = \sigma \downarrow$}. Therefore, with all market parameters being unchanged but increasing interest rates the manager has higher chances to yield a decent return for the investor, whence she requires a higher management fee for her job. Simultaneously, with high enough interest rates and good enough risk-return profile of the hedge-fund's risky investment opportunity, the investor tends to care less for the first-loss coverage guarantee, which is why $\hat{c}$ decreases.

Table \ref{tab:unc_pref_fee_structures_gamma} illustrates the influence of the market price of risk $\gamma$  on the preferred fee structures. We observe no trend for the preferred management fee or the preferred performance fee. Interestingly, the preferred first-loss coverage guarantee tends to decrease. So, the investor is willing to cut down on the first-loss coverage guarantee as the financial market offers a higher excess return per unit of risk.

Finally, we also investigate numerically the fund's Sharpe ratio for the manager's optimal strategy corresponding to the preferred first-loss fee structure. In Table \ref{tab_SR_optimal_and_benchmark} we compare the fund's Sharpe ratio as well as the parties' expected utilities for the optimal terminal fund value and the terminal values of different funds that follow specific constant-mix strategies. We write $V_T^{(1 - \pi_{CM}, \pi_{CM})}, \pi_{CM} \in \{ 0.25, 0.5, 0.75, 1\}$ for the fund's terminal value if the manager follows a constant mix strategy, where the constant proportion $\pi_{CM}$ of the budget is invested in the risky asset. In this table, we also use the notation $SR(V_T^{\cdot})$ to emphasize that the fund's Sharpe ratio explicitly depends on the fund's terminal value but does not explicitly depend on the preferred fee structure. In contrast, the expected utilities of the parties depend on both the fee scheme and the terminal fund value. We observe that the constant-mix strategies yield a slightly higher Sharpe ratio than the manager's optimal (in the sense of Problem \eqref{eq:nonc_optim_wealth_problem_man}) trading strategy. However, as anticipated, due to the distribution of the fund's terminal value between the parties, the expected utility of the manager is much lower for the constant-mix strategies than that of a manager following the optimal trading strategy. Interestingly, when the manager executes a constant-mix strategy, the investor has a slightly lower or a comparable expected utility compared to the investor's expected utility when the manager follows her optimal trading strategy.

\begin{table}[!ht]
	\begin{center}
		\caption{Comparison of the fund's Sharpe ratios as well as parties value functions for the optimal trading strategy and constant mix trading strategies, given the preferred fee structure choice}\label{tab_SR_optimal_and_benchmark}
\begin{tabular}{l|l|ccccc}
\toprule
Param. specification & Quantity &$ V_T^{*}$ & $V_T^{(0, 1)}$ & $V_T^{(0.25, 0.75)}$ & $V_T^{(0.5, 0.5)}$ & $V_T^{(0.75, 0.25)}$\\ 
\midrule
$b_M = b_I = 0.65$ &  $SR(V_T^{\cdot})$ & $34.06\%$ & {\color{green}$38.15\%$} & {\color{green}$38.73\%$} & {\color{green}$39.30\%$} & {\color{green}$39.97\%$} \\ 

$(\hat{m}, \hat{\alpha}, \hat{c}) = $ &$\phi_M(\hat{m}, \hat{\alpha}, \hat{c}|V_T^{\cdot})$ &  $2.1118$ & {\color{red}$1.6349$} & {\color{red}$1.6899$} & {\color{red}$1.768$} & {\color{red}$1.8531$} \\

$(5.0\%, 35.5\%, 26.0\%)$ & $\phi_I(\hat{m}, \hat{\alpha}, \hat{c}|V_T^{\cdot})$ & $3.1897$ & {\color{red}$3.1295$} & {\color{red} $3.1372$} & {\color{red}$3.138$} & {\color{red}$3.1339$} \\ 
\midrule
$b_M = b_I = 2.5$ & $SR(V_T^{\cdot})$ & $37.80\%$ & {\color{green}$38.15\%$} & {\color{green}$38.72\%$} & {\color{green}$39.30\%$} & {\color{green}$39.97\%$} \\ 

$(\hat{m}, \hat{\alpha}, \hat{c}) = $ & $\phi_M(\hat{m}, \hat{\alpha}, \hat{c}|V_T^{\cdot})$ & $ -3.5558$ & {\color{red}$-18.988$} & {\color{red}$-12.513$} & {\color{red}$-6.7601$} & {\color{red}$-4.4256$} \\ 

$(4.8\%, 50.0\%, 30\%)$ & $\phi_I(\hat{m}, \hat{\alpha}, \hat{c}|V_T^{\cdot})$ & $-0.4492$ & {\color{red}$ -0.4498$} & {\color{green}$-0.44625$} &  {\color{green}$-0.4467$}& {\color{green}$ -0.4488$} \\ 
\bottomrule 
\end{tabular}
	\end{center}
\end{table}
 
The hedge fund's Sharpe ratio predominantly increases when the manager or the investor becomes more risk-averse. In each considered combinations of the risk-aversion parameters in Table \ref{tab:unc_pref_fee_structures_bM_bI}, it is greater than the Sharpe ratio of the hedge fund with the traditional fee structure $(0\%,20\%, 0\%)$ by about $25$ percentage points on average and greater than the Sharpe ratio originating from the fee structure  $(0\%,30\%, 10\%)$ by around $12$ percentage points on average. The volatility of the hedge fund with the preferred first-loss scheme is decreasing in both $b_M$ and $b_I$. On average, the volatility of the hedge fund with the preferred first-loss fee structure is about $50\%$ lower than the volatility of the hedge fund with the fee structure $(0\%,20\%, 0\%)$ and around $17\%$ times lower than the volatility of the hedge fund with the fee structure $(0\%,30\%, 10\%)$. So the preferred first-loss schemes significantly decrease the fund's risk and increase the fund's Sharpe ratio.

Pareto optimality of the preferred fee structures does, however, not ensure that both the manager and the investor are better off when switching a traditional fee structure to a preferred first-loss scheme. In fact, in all cases from Table \ref{tab:unc_pref_fee_structures_bM_bI}  the investor's expected utility function is considerably higher than it is for a traditional fee structure $(0\%, 20\%, 0\%)$, whereas the manager's expected utility is slightly worse than it is for a traditional scheme  $(0\%, 20\%, 0\%)$. If the hedge-fund manager wants to ensure that she only offers first-loss fee structures yielding an expected utility that is not worse than that at the traditional fee structure, we recommend to choose the constrained preferred fee structure in the following way:
\begin{equation}\label{eq:eq_fee_opt_problem_2}
  \begin{aligned}
    \max_{m, \alpha, c} & \,\,\, SR^*(m, \alpha, c); \\
    \text{s.t.} &\,\,\, (m, \alpha, c) \in \mathcal{P}_{FBPO}^{\tilde \phi_M};\\
  \end{aligned}
\end{equation}
where  $\mathcal{P}_{FBPO}^{\tilde \phi_M} := \left\{ (m,\alpha, c) \in \mathcal{P}_{FBPO}: \phi_M (m,\alpha, c)  \geq  \tilde \phi_M \right\}$ and $\tilde \phi_M := \phi_M(\tilde m, \tilde \alpha, 0\%)$ for a fixed management fee $\tilde m$ and a fixed performance fee $\tilde \alpha$ that the manager charged in the traditional scheme.

Our numerical studies show that the preferred fee structures in the optimization \eqref{eq:eq_fee_opt_problem_2} usually have a slightly lower management fee as well as first-loss coverage guarantee and higher performance fee than the corresponding components of the preferred fee structures in the optimization \eqref{eq:eq_fee_opt_problem}.

\section{Conclusion}\label{sec:conclusion}
In the present paper, we study in an expected utility framework the optimal choice of first-loss hedge-fund fee structures. We  focus on the fee arrangements for which the investor's assets and the manager's deposit account are segregated, which is different from the case when the parties' assets are commingled.
First, we derive the fund's optimal terminal value combining the martingale approach and the concavification technique. Second, we calculate and analyze Pareto optimal fee structures. 
We choose the model parameters that account for the historic popularity of the $2\% \& 20\%$ arrangement in the hedge-fund sector and are consistent with the relevant literature as well as the current interest-rate environment. For the calibrated model, we find that the Pareto optimal fee structure that is closest to the $2\% \& 20\%$ scheme has a $0\%$ management fee and a $20.3\%$ performance fee. This is a possible explanation to the current trend of decreasing management fees in hedge funds that still use the traditional scheme.

For determining a single first-loss fee structure that can be considered fair by both the hedge fund manager and the investor, we follow \cite{EscobarAnel2018} and maximize the fund's Sharpe ratio on the set of Pareto optimal first-loss fee structures. We find that the traditional fee they recommend ($0\%$ management fee and $30.7\%$ performance fee) is not Pareto optimal in the presence of the first-loss coverage guarantee.  The fee structure  recommended in \cite{He2018} ($30\%$ performance fee and $10\%$ first-loss coverage guarantee) is not Pareto optimal either. The reason for that is that we consider a different type of the first-loss fee arrangement, as mentioned before. It is beyond the scope of this paper, but would be interesting to see whether this is different for their hedge-fund model. In our numerical studies, the preferred fee structures have the management fee usually close to $5\%$, the performance fee mainly in the range $30\% - 50\%$ and  the first-loss coverage in the range $15\% - 30\%$. 
However, if this fee structure is replaced by the traditional $2\% \& 20\%$ fee arrangement, the manager is worse off in terms of her expected utility, whereas the investor is much better off.
Hence, we also consider only those Pareto optimal fee structures that yield both parties at least the utility she has with a reasonable traditional fee structure. We get under this condition that the preferred management fee is $5\%$, performance fee is about $48\%$, and first-loss coverage guarantee is around $24\%$. All these findings shed light on the first-loss scheme that maturing hedge funds with first-loss compensation might be using in the future. Clearly, our model setting simplifies the real world, e.g. the price process follows a geometric Brownian motion, no transaction costs, etc. Therefore, it would be interesting to investigate the preferred fee structures in a more realistic setting, which is subject to future research.

The sensitivity analysis shows that the more risk-averse the investor becomes, the higher the preferred performance fee and the first-loss coverage guarantee should be.  As the manager's risk aversion increases, the preferred performance fee and the first-loss coverage guarantee tend to decrease. This is different to the findings in \cite{He2018}. However, the authors assume that the parties are loss averse and consider hedge funds with commingled assets of  investors and managers. The preferred first-loss coverage guarantee is decreasing in the market price of risk and increasing in the interest rates. Finally, we also found that derived preferred fee structures substantially decrease hedge fund's volatility in comparison to the traditional schemes as well as other reported first-loss fee structures.

\appendix

\section{Proofs of Main Results}\label{app:proofs}
\begin{proof}[Proof of Lemma \ref{lem:concavification_3_pieces}]
We show how to construct for $U_M$ its concave envelope $u_M$, the uniqueness of $\theta_1$ will follow. For $u_M$, four cases are possible, which are illustrated in Figure \ref{fig:concave_envelope_cases}.
Note that $U_M$ has the following properties:
\begin{enumerate}
  \item[(i)] $U_{M,1}:(0, \Theta_1) \rightarrow \mathbb{R}$ is constant\footnote{The proof is also valid, if (i) is less restrictive, i.e. $U_{M,1}:(0, \Theta_1) \rightarrow \mathbb{R}$ is convex and non-decreasing};
  \item[(ii)] $U_{M,2}:(\Theta_1, \Theta_2) \rightarrow \mathbb{R}$ is strictly increasing, strictly concave, continuously differentiable;
  \item [(iii)] $U_{M,3}:(\Theta_2, +\infty) \rightarrow \mathbb{R}$ is strictly increasing, strictly concave and continuously differentiable with $\lim_{v \uparrow +\infty} U_{M,3}'(v) = 0$;
  \item[(iv)] $U_{M,2}'(\Theta_2-) \geq U_{M,3}'(\Theta_2+)$;
  \item[(v)] $U_{M,i}(\Theta_i-) = U_{M,i+1}(\Theta_i+)$ for $i \in \{ 1,2\}$.
\end{enumerate}

\begin{figure}[!ht]
\begin{subfigure}{.5\textwidth}
  \centering
  \includegraphics[width=.95\linewidth]{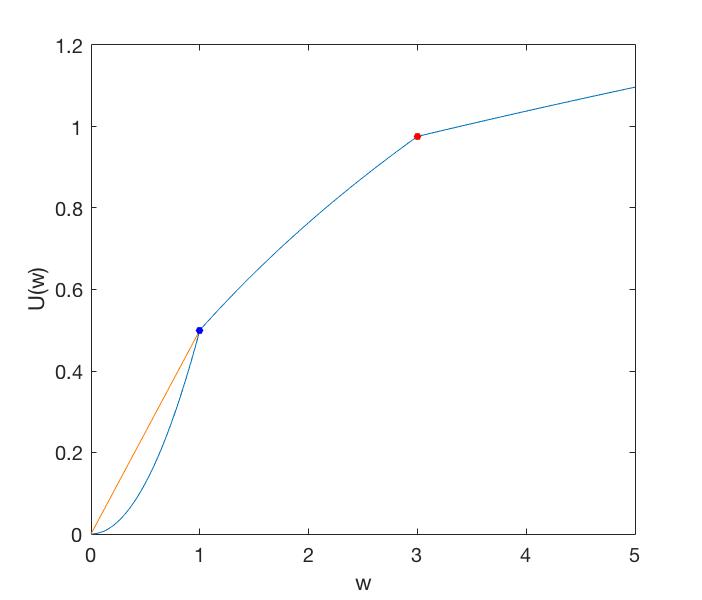}
  \caption{Case 1}
  \label{sfig:3p_concave_env_1}
\end{subfigure}%
\begin{subfigure}{.5\textwidth}
  \centering
  \includegraphics[width=.95\linewidth]{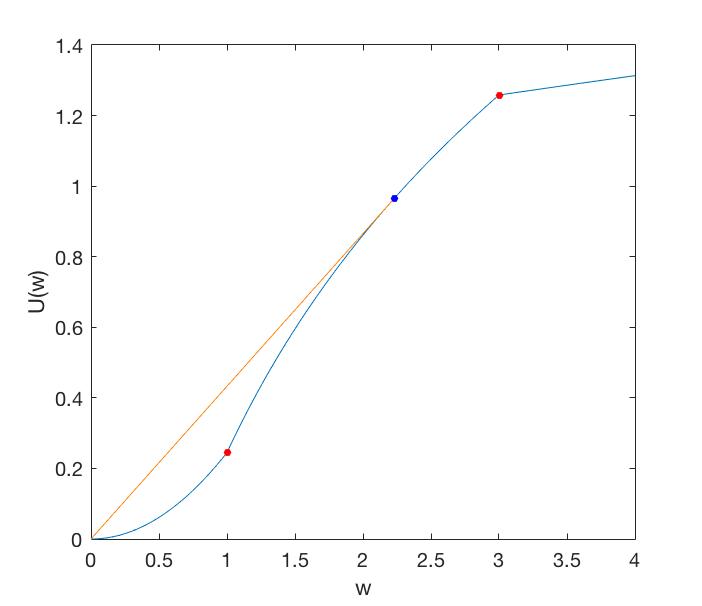}
  \caption{Case 2}
  \label{sfig:3p_concave_env_2}
\end{subfigure}
\\
\begin{subfigure}{.5\textwidth}
  \centering
  \includegraphics[width=.95\linewidth]{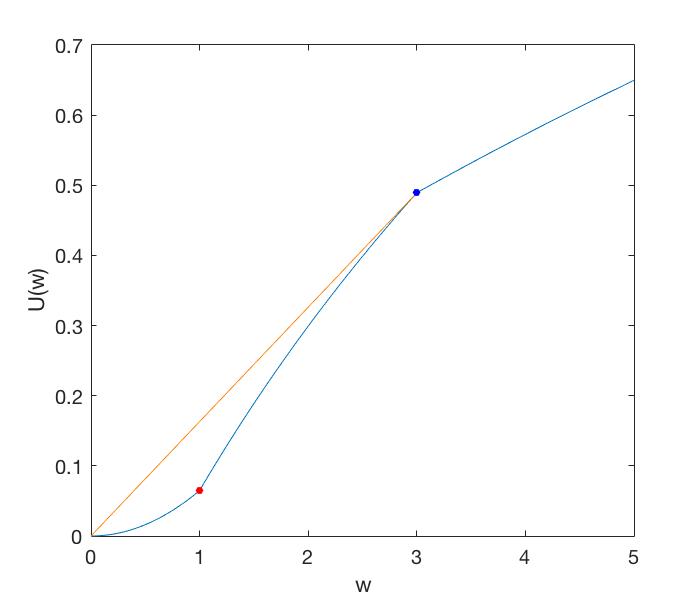}
  \caption{Case 3}
  \label{sfig:3p_concave_env_3}
\end{subfigure}%
\begin{subfigure}{.5\textwidth}
  \centering
  \includegraphics[width=.95\linewidth]{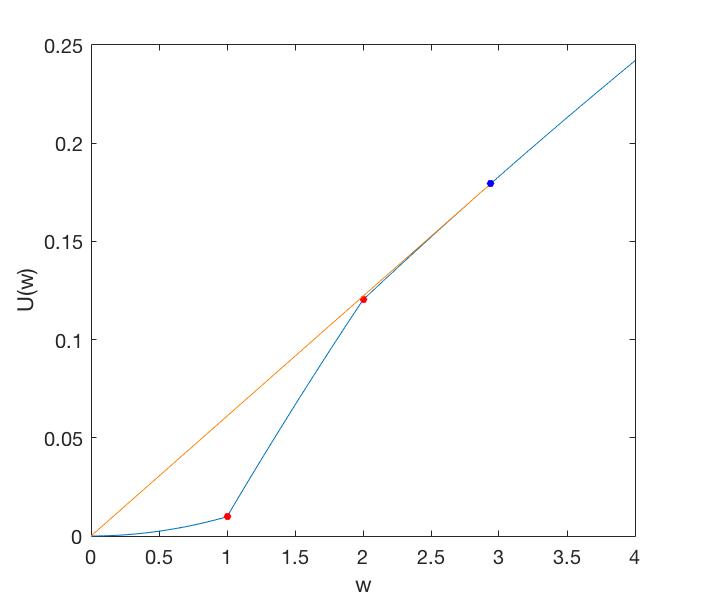}
  \caption{Case 4}
  \label{sfig:3p_concave_env_4}
\end{subfigure}
\caption{Utility functions and their concave envelopes}
\label{fig:concave_envelope_cases}
\end{figure}

Denote $s(v) := (U_M(v) - U_M(0))/v$ the slope of a line that passes through the points $(0, U_M(0))$ and $(v, U_M(v))$. Since $U_M$ is nondecreasing, $s(v) \geq 0\,\forall v > 0$. 

Due to properties (ii)-(v) $U_{M,2}'(\Theta_2-) \geq U_{M,3}'(\Theta_2+)$, $U_{M,2}$ and $U_{M,3}$ are strictly concave on $(\Theta_1; \Theta_2)$ and $(\Theta_2, +\infty)$ respectively, $U_{M,2}(\Theta_2-) = U_{M,3}(\Theta_2+)$, we conclude that $U_M(v)$ is strictly concave on $(\Theta_1, +\infty)$. 

Since $U_{M,1}(v)$ is convex (see property (i)), we get:
\begin{equation}\label{eq:convexity_U_1}
U_{M,1}(v) \leq U_{M,1}(0) + s(\Theta_1)v,\,\,\,\,\,\forall v \in [0, \Theta_1].
\end{equation}
Therefore, $\theta_1 \geq \Theta_1$. Next we partition the set of functions $U_M$, satisfying the assumptions of this lemma, into four subsets, depending on the values of $s(\Theta_1)\geq 0, s(\Theta_2) > 0$. We show that in each case there is a unique $\theta_1$ -- the change point between the linear part of the concave envelope and $U_M$.

Case 1: $s(\Theta_1) \in [U_{M,2}'(\Theta_1+), +\infty)$ and any $s(\Theta_2)$.

Since the linear function $g(v)=U_M(0) + s(\Theta_1)v$ is obviously concave on $[0, \Theta_1]$, $U_M(v)$ is strictly concave on $[\Theta_1, +\infty)$ and $g(\Theta_1-) = s(\Theta_1) \geq U_M(\Theta_1+) = U_{M,2}(\Theta_1+)$, we
conclude that the function
$$(U_M(0) + s(\Theta_1)v)\mathbbm{1}_{[0, \Theta_1)}(v) + U_M(v)\mathbbm{1}_{[\Theta_1, +\infty)}(v)$$
is concave.  The candidate for the concave envelope of $U_M$ is
\begin{equation*}
u_M(v) = \left\{
\begin{aligned}
&-\infty, && v < 0;\\
&U_M(0) + s(\Theta_1)v, && v \in [0, \Theta_1); \\
&U_{M,2}(v), && v \in [\Theta_1, \Theta_2); \\
&U_{M,3}(v), && v \in [\Theta_2, +\infty).
\end{aligned}\right.
\end{equation*}
Take any concave function $g \geq U_M$. We get that $g(v) \geq U_M(v) = u_M(v)$ for $v \in (-\infty, 0] \cup [\Theta_1, +\infty)$. Since $u_M$ is linear on $(0, \Theta_1)$ and $g$ is concave, we get for any $v = \lambda \Theta_1$ with $\lambda \in (0, 1)$:
\begin{equation*}
	\begin{aligned}
		g(v) & = g(\lambda \Theta_1 + (1 - \lambda) 0)  \geq \lambda g(\Theta_1) + (1 - \lambda) g(0) \geq \lambda U_M(\Theta_1) + (1 - \lambda) U_M(0) \\
		& = \lambda u_M(\Theta_1) + (1 - \lambda) u_M(0) = u_M(\lambda \Theta_1 + (1 - \lambda) 0) = u_M(v).
	\end{aligned}
\end{equation*}
Then according to Definition \ref{def:concave_envelope}, $u_M$ is the concave envelope of $U_M$.  Writing it down in the form (\ref{eq:utility_piecewise}) is straightforward setting $\theta_1 = \Theta_1$ and $\theta_2 = \Theta_2$. Figure \ref{sfig:3p_concave_env_1} provides graphical illustration to this case. Red markers stand for the change points of $U_M$, i.e. $\Theta_1$, $\Theta_2$. The blue marker corresponds to the point $\theta_1$.

Case 2: $s(\Theta_1) \in [0, U_{M,2}'(\Theta_1+))$ and $s(\Theta_2) \in (U_{M,2}'(\Theta_2-), +\infty)$.

Note that:
\begin{equation}\label{eq:boundary_conditopn_a2}
\begin{aligned}
s(\Theta_1) < U_{M,2}'(\Theta_1+)  \Leftrightarrow U_{M,2}(\Theta_1) - U_{M,1}(0) - U_{M,2}'(\Theta_1+)\Theta_1 < 0; \\
s(\Theta_1) > U_{M,2}'(\Theta_2-) \Leftrightarrow U_{M,2}(\Theta_2) - U_{M,1}(0) - U_{M,2}'(\Theta_2-)\Theta_2 > 0.
\end{aligned} 
\end{equation}
Using (\ref{eq:boundary_conditopn_a2}) and property (ii), it is easy to show that there exists a unique $\theta_1 \in (\Theta_1, \Theta_2)$, such that $s(\theta_1) = U_{M,2}'(\theta_1)$.


Now we show that $s(v)$ is strictly increasing on $(\Theta_1, \theta_1)$. Consider the derivative of $s(v)$ on $(\Theta_1, \Theta_2)$:
\begin{equation*}
s'(v) = \frac{U_{M,2}'(v)v - U_{M,2}(v) + U_{M,1}(0)}{v^2} = \frac{-a_2(v)}{v^2}.
\end{equation*}
Since $a_2(v)$ is strictly increasing on $(\Theta_1, \Theta_2)$ and $a_2(\theta_1) = 0$, we conclude that $a_2(v) < 0$ for $v \in (\Theta_1, \theta_1)$. Therefore, $s'(v) = \frac{-a_2(v)}{v^2} > 0$ for $v \in (\Theta_1, \theta_1)$, whence $s(\Theta_1) < s(\theta_1)$.

Using $s(\Theta_1) < s(\theta_1)$ and strict concavity and differentiability of $U_{M,2}$ on $(\Theta_1, \Theta_2)$ (see property (ii)), we obtain:
\begin{equation*}
\begin{aligned}
U_{M,1}(v) & \leq U_M(0) + s(\Theta_1)v < U_M(0) + s(\theta_1)v, \,\,\,\, \forall v \in [0,\Theta_1);\\
U_{M,2}(v) &< U_{M,2}(\theta_1) + \underbrace{U_{M,2}'(\theta_1)}_{=s(\theta_1)} (v - \theta_1)  \\
& =U_{M,2}(\theta_1) + \frac{U_{M,2}(\theta_1) - U_{M,1}(0)}{\theta_1}(v - \theta_1) \\
& =  U_{M,2}(\theta_1) + s(\theta_1)v - (U_{M,2}(\theta_1) - U_M(0)) \\
& = U_M(0) + s(\theta_1)v,   \,\, \forall v \in [\Theta_1, \theta_1).
\end{aligned}
\end{equation*}

Finally, one can check similarly to Case 1 that the function:
\begin{equation*}
u_M(v) = \left\{
\begin{aligned}
&-\infty, && v < 0;\\
&U_M(0) + s(\theta_1)v, && v \in [0, \theta_1); \\
&U_{M,2}(v), && v \in [\theta_1, \Theta_2); \\
&U_{M,3}(v), && v \in [\Theta_2, +\infty)
\end{aligned}\right.
\end{equation*}
is indeed the concave envelope of $U_M$ in the sense of Definition \ref{def:concave_envelope}. In terms of (\ref{eq:utility_piecewise}), we have a unique $\theta_1 \in (\Theta_1, \Theta_2)$ and can set $\theta_2 = \Theta_2$.
Figure \ref{sfig:3p_concave_env_2} corresponds to Case 2.

Case 3: $s(\Theta_1) \in [0, U_{M,2}'(\Theta_1+)],\, s(\Theta_2) \in[U_{M,3}'(\Theta_2+) ,U_{M,2}'(\Theta_2-)]$.
 This case is similar to Case 1 and is illustrated  in Figure \ref{sfig:3p_concave_env_3}. In terms of (\ref{eq:utility_piecewise}), we have $\theta_1 = \theta_2 = \Theta_2$.

Case 4: $s(\Theta_1) \in [0, U_{M,2}'(\Theta_1+)],\, s(\Theta_2) \in(0, U_{M,3}'(\Theta_2+)).$
This case is analogous to Case 2 and is illustrated  in Figure \ref{sfig:3p_concave_env_4}. In terms of (\ref{eq:utility_piecewise}), we have a unique $\theta_1 \in (\Theta_2, +\infty)$ and can set $\theta_2 = \theta_1$.

We have covered all possible cases, i.e. classified all possible values of $s(\Theta_1)\geq 0, s(\Theta_2) > 0$. Any $U_M$ that has properties (i)-(v) is related to exactly one of the four described cases.
\end{proof}

\begin{proof}[Proof of Theorem \ref{th:optim_w_linear_part_1}] 
1. The equation $h(y) - v_0 = 0$ has a unique solution $y^* \in (0, +\infty)$ due to the fact that under the integrability assumption (\ref{as:integrability_1}) the function $h(y) - v_0$ is continuous, strictly decreasing, $\lim_{y \uparrow +\infty} (h(y) - v_0)= -v_0 < 0$ as well as $\lim_{y \downarrow 0} (h(y) - v_0) = +\infty$.\\
2. Denote $V^*(T) := v^*(y^*, \tilde{Z}(T))$. Take any admissible solution $\bar{V}(T)$ of problem (\ref{eq:conc_optim_wealth_problem_man}). It must satisfy the budget constraint $\mathbb{E}[\tilde{Z}(T)\bar{V}(T)] \leq v_0$. Then:
 \begin{flalign*}
\mathbb{E}[u_M(V^*(T))] &- \mathbb{E}[u_M(\bar{V}(T))] =  \mathbb{E}[u_M(V^*(T))] - y^*v_0  - \mathbb{E}[u_M(\bar{V}(T)] + y^* v_0\\
& \geq \mathbb{E}[u_M(V^*(T))] - y^*v_0 - \mathbb{E}[u_M(\bar{V}(T))] + y^*\mathbb{E}[\tilde{Z}(T)\bar{V}(T)] \\
& = (\mathbb{E}[u_M(V^*(T))] - y^*\mathbb{E}[\tilde{Z}(T)V^*(T)]) - (\mathbb{E}[u_M(\bar{V}(T))]\\
&\,\,\,\,\,\, - y^*\mathbb{E}[\tilde{Z}(T)\bar{V}(T)])  \\
& = \mathbb{E}[(u_M(V^*(T)) - y^*\tilde{Z}(T)V^*(T)) - (u_M(\bar{V}(T)) \\
& \,\,\,\,\,\,- y^*\tilde{Z}(T)\bar{V}(T))]  \stackrel{Lem.\ref{lem:pointwise_optim_sol_linear_1}}{\geq} 0.
\end{flalign*}
Hence, $V_T^*$ is optimal.  It is $\Q$-a.s. unique, since $\Q(\tilde{Z}_T = u_{M,1}'(\theta_1-) / y^*) = 0$.\\
3. Observe that the initial utility function $U_M(v)$ and its concave envelope $u_M(v)$ satisfy the following properties:
\begin{equation}\label{eq:solutions_relations}
\begin{aligned}
\{ v \in \mathbb{R}: U_M(v) < u_M(v)\} &= (0,\theta_1);\\
 \{ v \in \mathbb{R}: U_M(v) = u_M(v)\} &= \mathbb{R}\setminus(0,\theta_1).
\end{aligned}
\end{equation}
Let $V_T^*$ be the optimal solution of Problem $\eqref{eq:conc_optim_wealth_problem_man}$. Since the feasible regions in Problem $\eqref{eq:nonc_optim_wealth_problem_man}$ and Problem $\eqref{eq:conc_optim_wealth_problem_man}$ coincide, $V_T^*$ is feasible in the former (original, non-concave) problem. Note that $\mathbb{P}(V_T^* \in \{ 0 \} \cup [\theta_1, +\infty)) = 1$ due to the definition of $\theta_1$.  Denote $V_T$ any other feasible wealth in Problem $\eqref{eq:nonc_optim_wealth_problem_man}$ such that $\mathbb{Q}(V_T=V_T^*)\neq 1$. Then:

\begin{equation*}
\begin{aligned}
\mathbb{E}[U_M(V_T^*)]  - \mathbb{E}[U_M(V_T)] \stackrel{(\ref{eq:solutions_relations})}{=} & \underbrace{\mathbb{E}\left[\left(u_M(V_T^*) - u_M(V_T\right)\mathbbm{1}_{\{ V_T^* \not \in (0,\theta_1) \}}\right]}_{\geq0 \text{, due to optimality of } V_T^* \text{ in Problem (\ref{eq:conc_optim_wealth_problem_man}) } } \\
& + \underbrace{\mathbb{E}\left[\left(U_M(V_T^*) - U_M(V_T\right)\mathbbm{1}_{\{ V_T^* \in (0,\theta_1) \}}\right]}_{=0,\text{ since }\mathbb{Q}(V_T^* \in (0,\theta_1)) = 0}  \geq 0.
\end{aligned}
\end{equation*}
Hence, $V_T^*$ is the $\Q$-a.s. unique optimal solution of Problem (\ref{eq:nonc_optim_wealth_problem_man}).
\end{proof}

\begin{proof}[Proof of Proposition \ref{prop:nonlin_opt_solution_existence}]
From Theorem \ref{th:optim_w_linear_part_1}, we know that $V_T^* = v(y^*, \tilde{Z}_T)$ for $y^*$ such that $\E{V_T^*\tilde{Z}_T}=v_0$ and $v(y, \tilde{z})$ solving \eqref{eq:point_optim_problem}.  The function $v(y, \tilde{z})$, derived in the supplementary Lemma \ref{lem:pointwise_optim_sol_linear_1}, \ref{app:aux_proofs}, is continuous w.r.t. $(m, \alpha, c)$. Hence, the function $\phi_M(m, \alpha, c)$ is continuous w.r.t. $(m, \alpha, c) \in \mathcal{P}$ as a superposition of continuous functions.

Fix $\displaystyle \phi_{min} \in \left[\min_{(m, \alpha, c) \in \mathcal{P}}\phi_M(m, \alpha, c),\,\max_{(m, \alpha, c) \in \mathcal{P}}\phi_M(m, \alpha, c)\right]$. Then the set \\$\mathcal{R} = \left\{ (m, \alpha, c):\phi_M(m, \alpha, c) \geq \phi_{min} \right\}$ is non-empty due to the choice of $\phi_{min}$ and closed due to the continuity of $\phi_M(m, \alpha, c)$. The set $\mathcal{P}$ is non-empty, closed and bounded. Hence, $\mathcal{R} \cap \mathcal{P}$ is a non-empty, closed and bounded set.

The function $\phi_I(m, \alpha, c)$ is also continuous for $(m, \alpha, c) \in \mathcal{P}$ as a superposition of continuous functions. Therefore, by the Weierstrass extreme value theorem, there exists $(m^*, \alpha^*, c^*) \in \mathcal{R} \cap \mathcal{P}$ such that $\phi_I(m^*, \alpha^*, c^*) \geq \phi_I(m, \alpha, c)\, \forall (m, \alpha, c) \in \mathcal{R} \cap \mathcal{P}$.
%
%
\end{proof}

\begin{proof}[Proof of Corollary \ref{cor:fl3_optimal_terminal_vaule}]
This corollary is a direct application of  Lemma \ref{lem:concavification_3_pieces} (the construction of the concave envelope of the manager's utility function) and Theorem \ref{th:optim_w_linear_part_1}.

Recall that $\Theta_1 = (1+m - c) v_0$,  $\Theta_2 = (1+m)v_0$, $s(v) = (U_M(v) - U_M(0))/v$. Then:
\begin{equation}\label{eq:utility_derivatives_at_kinks}
\begin{aligned}
U_M'(\Theta_1+) &= \lim_{v \downarrow \Theta_1} U_{M,2}'(v) = \lim_{v \downarrow \Theta_1} (v - v_0 + a_M)^{-b_M} = ((m - c)v_0 + a_M)^{-b_M};\\
U_M'(\Theta_2-) &=  \lim_{v \uparrow \Theta_2} U_{M,2}'(v) = \lim_{v \uparrow \Theta_2} (v - v_0 + a_M)^{-b_M} = (mv_0 + a_M)^{-b_M};\\
U_M'(\Theta_2+) &=  \lim_{v \downarrow \Theta_2} U_{M,3}'(v) = \lim_{v \downarrow \Theta_2} (\alpha v + mv_0 - \alpha(1+m)v_0 + a_M)^{-b_M}\alpha = \alpha(mv_0 + a_M)^{-b_M}.
\end{aligned}
\end{equation}

Since it holds that $s(\Theta_1) = 0 < U_{M,2}'(\Theta_1+)$, one can easily verify that Case 2, Case 3 and Case 4 from Lemma \ref{lem:concavification_3_pieces} correspond to $\mathcal{P}_A$ (Case A), $\mathcal{P}_B$ (Case B) and $\mathcal{P}_C$ (Case C) respectively, where $\mathcal{P}_{X}$, $X \in \{ A, B, C\}$, is defined in (\ref{eq:parametric_regions}).

Consider Case A, $(m,\alpha, c) \in \mathcal{P}_A$. According to Lemma \ref{lem:concavification_3_pieces}, there exists a unique $\theta_1 > \Theta_2$ solving:
\begin{equation*}
\begin{aligned}
&s(v) = U_M'(v) \Leftrightarrow \frac{(\alpha v + (m -  \alpha(1+m))v_0 + a_M)^{1-b_M} - (v_0(m - c) + a_M)^{1-b_M}}{(1 - b_M)v} = U_{M,3}'(v) \\
&  \Leftrightarrow (\alpha v + (m -  \alpha(1+m))v_0 + a_M)^{-b_M} (b_M \alpha v + (m -  \alpha(1+m))v_0 + a_M) =   (v_0(m - c) + a_M)^{1-b_M}.
\end{aligned}
\end{equation*}
The concave envelope of $U_M$ in this case has the following form:
\begin{equation}\label{eq:concave_envelope_A}
 u_M(V_T) = \left\{
\begin{aligned}
&-\infty,   &&V_T < 0;\\
&\frac{1}{1 - b_M}(v_0(m - c) + a_M)^{1 - b_M}+ s(\theta_1)V_T, &&V_T \in [0,\theta_1); \\
&\underbrace{\frac{1}{1 - b_M}(\alpha V_T + (m -  \alpha(1+m))v_0 + a_M)^{1-b_M}}_{= U_M(V_T) = U_{M,3}(V_T) \,\,\, \forall V_T \in [\theta_1,+\infty)}, &&V_T \in [\theta_1,+\infty).
\end{aligned}\right.
\end{equation}



Denote by $I_2(\cdot), I_3(\cdot)$ the inverse functions of the marginal utilities $U_{M,2}'(\cdot), U_{M,3}'(\cdot)$:
\begin{equation*}
\begin{aligned}
 & I_2(v) = v^{-\frac{1}{b_M}} + v_0 - a_M,\, I_3(v) = \alpha^{\frac{1}{b_M} - 1}v^{-\frac{1}{b_M}} +  (1+m - \alpha^{-1}m)v_0 - \alpha^{-1}a_M.
\end{aligned}
\end{equation*}

Then for $v \in [0, +\infty)$ the concave envelope $u_M(v)$ has the form of \eqref{eq:utility_piecewise} for $u_{M, 2}(\cdot) = u_{M, 3}(\cdot)$ and  $\theta_2 = \theta_1$. Using that  $u_{M, 2}(\cdot) = u_{M, 3}(\cdot)$, $u_{M, 1}'(\theta_1-) = u_{M, 2}'(\theta_1+) = s(\theta_1) = \alpha(\alpha \theta_1 + (m - \alpha(1+m))v_0 +a_M)^{-b_M}$ and setting $\tilde{v}_1 = \theta_1$, we obtain by applying Lemma \ref{lem:pointwise_optim_sol_linear_1} that
\begin{equation*}
v^*(y, \tilde{z}) = \left( \alpha^{1/b_M - 1}(y\tilde{z})^{-1/b_M} +  (1+m - \alpha^{-1}m)v_0 - \alpha^{-1}a_M \right)\mathbbm{1}_{\left(0,\, s(\theta_1)/y\right]}(\tilde{z})
\end{equation*}
solves $\max_{v\geq 0}\{ u_M(v) - y \cdot \tilde{z} \cdot v \}$.

Using the supplementary Lemma \ref{lem:expectation_Zk_ab} in \ref{app:aux_proofs} one can easily show that the integrability condition in Theorem \ref{th:optim_w_linear_part_1} holds for  $\forall y > 0$.
Therefore, we may apply Theorem \ref{th:optim_w_linear_part_1} and conclude that:
\begin{equation*}
\begin{aligned}
V^*_T = v^*(y^*, \tilde{Z}_T) =& \left( \alpha^{1/b_M - 1}(y^*\tilde{Z}_T)^{-1/b_M} +  (1+m - \alpha^{-1}m)v_0 - \alpha^{-1}a_M \right) \mathbbm{1}_{\left\{\tilde{Z}_T \in \left(0,\, s(\theta_1^A)/y^*\right]\right\}},
\end{aligned}
\end{equation*}
where $y^* \in (0, +\infty)$ is the unique solution of the equation $\mathbb{E}\left[\tilde{Z}_Tv^*(y, \tilde{Z}_T)\right] =v_0$ and $\theta_1^A := \theta_1$ to emphasize the correspondence of this concavification to Case A.

Case B and Case C are proven analogously to Case A.
\end{proof}

\begin{proof}[Proof of Proposition \ref{prop:value_fct_manager}]
In the proof we mainly use  Corollary \ref{cor:fl3_optimal_terminal_vaule}.

Case A: $(m,\alpha, c) \in \mathcal{P}_A$. Denote $E_1 = \left(0, s(\theta_1^A)/y^*\right]$. According to (\ref{eq:case_A_optimal_V})
\begin{equation*}
 V_T^*= \left( \alpha^{1/b_M - 1}(y^*\tilde{Z}_T)^{-1/b_M} + (1+m - \alpha^{-1} m)v_0 - \alpha^{-1} a_M\right)\indV{\tilde{Z}_T \in E_1}.
\end{equation*}

Using that $s(\theta_1^A) = \alpha(\alpha \theta_1^A + (m - \alpha(1+m))v_0 +a_M)^{-b_M}$ and the fact that $V_T^*=v^*(y^*, \tilde{Z}_T)$ is a non-increasing function of $\tilde{Z}_T$, one can easily verify that: 
\begin{equation}\label{eq:case_A_V_regions}
  \begin{aligned}
    V_T^*(\omega) & \geq \theta_1^A > (1+m)v_0 & & \forall \omega \in \left\{ \omega \in \Omega : \tilde{Z}_T(\omega) \in E_1 \right\};\\
    V_T^*(\omega) &= 0 & &\forall \omega \in  \left\{ \omega \in \Omega : \tilde{Z}_T(\omega) \not \in E_1 \right\}.
  \end{aligned}
\end{equation}
Therefore, it holds:
\begin{equation}\label{eq:u_U_equality}
\begin{aligned}
 \mathbb{Q}( \omega \in \Omega :& V_T^*(\omega) \in \left\{ v \in \mathbb{R}: u_M(v) \neq U_M(v) \right\}) = \mathbb{Q}\left( \omega \in \Omega : V_T^*(\omega) \in (0, \theta_1^A) \right) = 0.
\end{aligned}
\end{equation}
Then: 
 \begin{align*}
    \phi_M (m, \alpha, c) &\,\,\,\, = \,\,\,\,\,  \E{U_M(V^*_T)} \stackrel{(\ref{eq:u_U_equality})}{=} \E{u_M(V^*_T)}\\
    & \stackrel[V_T^*\geq 0]{(\ref{eq:concave_envelope_A})}{=}  \E{\left( U_M(0) + s(\theta_1^A)V_T^*\right) \indV{V_T^* \in [0, \theta_1^A)}} \\
    &\qquad\,\,\,+ \mathbb{E}\Bigl[(1 - b_M)^{-1}(\alpha V_T^* + (m -  \alpha(1+m))v_0 + a_M)^{1-b_M}  \\
    &\qquad\,\,\, \cdot \indV{V_T^* \in[\theta_1^A, +\infty)} \Bigr]\\
    & \stackrel[(\ref{eq:case_A_optimal_V})]{(\ref{eq:case_A_V_regions})}{=} \mathbb{E}\Bigl[ (U_M(0) + s(\theta_1^A) \cdot 0) \indV{Z_T \notin E_1}\Bigr] + (1 - b_M)^{-1} \\
    & \qquad\,\,\, \cdot \mathbb{E}\biggl[\Bigl(  \alpha \left( \alpha^{1/b_M - 1} (y^*\tilde{Z}_T)^{-1/b_M}  + (1+m - \alpha^{-1} m)v_0 \right.\\
    &\qquad\,\,\, \left.  - \alpha^{-1} a_M\right)+ (m -  \alpha(1+m))v_0 + a_M \Bigr)^{1-b_M} \indV{\tilde{Z}_T\in E_1}\biggr]  \\
    &\,\,\,\, = \,\,\,\,  U_M(0)\E{\indV{\tilde{Z}_T \notin E_1}} +  (1 - b_M)^{-1} \left(y^*\right)^{1 - 1/b_M}\alpha^{1/b_M - 1}   \\
    &\qquad\,\,\, \cdot \E{\tilde{Z}_T^{1-1/b_M}\indV{\tilde{Z}_T \in E_1}} \\ 
    & \stackrel[\tilde{Z}_T > 0]{\text{\scriptsize Lem.\ref{lem:expectation_Zk_ab}}}{=}  U_M(0)(\Phi(d_2^A) - \Phi(-\infty)) + \left(y^*\right)^{1 - 1/b_M}\alpha^{1/b_M - 1}  \\
    &\qquad\,\,\, \cdot \ExpTwoPar{(1/b_M - 1)}{0.5(1/b_M-1)^2} \\
    & \qquad\,\,\,\cdot \PhiDifFin{d_1^A}{d_2^A}{+(1 - b_M^{-1})}\\
    &\qquad\,\,\, \cdot  (1 - b_M)^{-1},
 \end{align*}
where
\begin{equation*}
  \begin{aligned}
  d_1^A & = +\infty, && d_2^A & = \frac{\log \left( y^* /s(\theta_1^A) \right) - \left( r+ 0.5\gamma^2 \right)T}{\gamma\sqrt{T}}; 
  \end{aligned}
\end{equation*}

Case B and Case C follow similarly.
\end{proof}

\begin{proof}[Proof of Proposition \ref{prop:value_fct_investor}]
In the proof of this proposition we mainly use Corollary \ref{cor:fl3_optimal_terminal_vaule}.

Case A: $(m,\alpha, c) \in \mathcal{P}_A$. Denote $E_1 = \left(0, s(\theta_1^A)/y^*\right]$.
We obtain:
\begin{align*}
 \phi_I(m, \alpha, c)  &\,\,\,\, = \,\,\,\,\, \E{(1 - b_I)^{-1}(I(V_T^*) + a_I)^{1- b_I}} \\
 & \,\,\, \stackrel{(\ref{eq:terminal_wealth_investor})}{=} \,\,\,\,(1 - b_I)^{-1}\mathbb{E}\biggl[ \Bigl( (V_T^*+v_0(c-m)) \indV{V_T^* \in [0, (1 + m - c)v_0)}\\
 &  \qquad \,\,\,+  v_0 \indV{V_T^* \in [(1 + m - c)v_0, (1 + m)v_0)}  + \Bigl(V_T^* - mv_0 \\
 &  \qquad \,\,\, - \alpha(V_T^* - (1+m)v_0)\Bigr) \indV{V_T^* \in [(1 + m)v_0, +\infty)} + a_I\Bigr)^{1-b_I}\biggr]\\
 & \stackrel[(\ref{eq:case_A_V_regions})]{(\ref{eq:case_A_optimal_V})}{=} (1 - b_I)^{-1}\mathbb{E}\biggl[ \Bigl( (0+v_0(c-m))\indV{\tilde{Z}_T\notin E_1}+ v_0 \mathbbm{1}_{\varnothing} + \Bigl( (1 - \alpha)\\
 &\qquad \,\,\, \cdot  \left( \alpha^{1/b_M - 1}(y^*\tilde{Z}_T)^{-1/b_M} + (1+m - \alpha^{-1} m)v_0 - \alpha^{-1} a_M\right) \\
 &\qquad \,\,\,  + (\alpha(1+m) - m) v_0 \Bigr)\indV{\tilde{Z}_T \in E_1} + a_I \Bigr)^{1 - b_I} \biggr]\\
 & \,\,\,\, = \,\,\,\,\, (1 - b_I)^{-1} \E{ \left( v_0(c - m) + a_I \right)^{1 - b_I}\indV{\tilde{Z}_T\notin E_1}} + (1 - b_I)^{-1} \\
 & \qquad \,\,\, \cdot  \mathbb{E}\biggl[ \Bigl( (1 - \alpha)\alpha^{1/b_M - 1}(y^*)^{-1/b_M}\tilde{Z}_T^{-1/b_M} \\
 & \qquad \,\,\, + (1+m - \alpha^{-1}m)v_0 - \alpha^{-1} a_M  - (\alpha(1+m) - m)v_0 \\
 &  \qquad \,\,\,  + a_M + (\alpha(1+m) - m)v_0 + a_I \Bigr)^{1 - b_I}\indV{\tilde{Z}_T\in E_1} \biggr] \\
 &  \,\,\,\, = \,\,\,\,\, (1 - b_I)^{-1}\left( v_0(c - m) + a_I \right)^{1 - b_I} \E{ \indV{\tilde{Z}_T\notin E_1}} + (1 - b_I)^{-1}\\
 & \qquad \,\,\, \cdot \mathbb{E}\biggl[ \Bigl( (1 - \alpha)\alpha^{1/b_M - 1} (y^*)^{-1/b_M}\tilde{Z}_T^{-1/b_M} + (1+m - \alpha^{-1}m)v_0 \\
 & \qquad \,\,\, + a_M(1 - \alpha^{-1}) + a_I\Bigr)^{1 - b_I} \indV{\tilde{Z}_T\in E_1} \biggr]\\
 & \stackrel[\tilde{Z}_T > 0]{\text{\scriptsize Lem.\ref{lem:expectation_Zk_ab}}}{=}  (1 - b_I)^{-1}\left( v_0(c - m) + a_I \right)^{1 - b_I} \left( \Phi(d_2^A)  - \Phi(-\infty)\right) \\
 & \qquad \,\,\, + (1 - b_I)^{-1} \mathbb{E} \biggl[ \Bigl( (1 - \alpha)\alpha^{1/b_M - 1}(y^*)^{-1/b_M}\tilde{Z}_T^{-1/b_M}  \\
 &  \qquad \,\,\,  + (1+m - \alpha^{-1}m)v_0 + a_M(1 - \alpha^{-1}) + a_I\Bigr)^{1 - b_I} \indV{\tilde{Z}_T\in E_1} \biggr]\\
 &  \,\,\,\, = \,\,\,\,\, (1 - b_I)^{-1}\left( v_0(c - m) + a_I \right)^{1 - b_I} \Phi(d_2^A) \\
 & \qquad \,\,\, + (1 - b_I)^{-1}\E{\left(k\tilde{Z}_T^{-1/b_M} + l\right)^{1 - b_I}\indV{\tilde{Z}_T\in E_1}},
\end{align*}
where $k = (1 - \alpha)\alpha^{1/b_M - 1}(y^*)^{-1/b_M}$ and $l = (1+m - \alpha^{-1}m)v_0  + a_M(1 - \alpha^{-1}) + a_I$, and $d_2^A$ is defined in Proposition \ref{prop:value_fct_manager}.

The derivation of $\phi_I(m, \alpha, c)$ for $(m,\alpha, c) \in \mathcal{P}_B$ and for $(m, \alpha, c) \in \mathcal{P}_C$ is done similarly to Case A.
\end{proof}

%
%

\section{Supplementary materials}\label{app:aux_proofs}

\begin{taggedlemma}{SL.1}\label{lem:expectation_Zk_ab}
Let $\tilde{Z}_T$ be the state price density process at time $T$, $a \in \mathbb{R}_{++}, b \in \mathbb{R}_{++}$, such that $a < b$. Then for any $k \in \mathbb{R}$ it holds:
\begin{equation*}
\begin{aligned}
 \mathbb{E}\left[ \tilde{Z}^k_T \mathbbm{1}_{\{ a<\tilde{Z}_T < b \}}\bigr| \mathcal{F}_t\right] =& \tilde{Z}^k_t e^{-k\left(r+\frac{\gamma^2}{2}\right)(T-t) + \frac{k^2\gamma^2}{2}(T-t)}
 \cdot \left(\Phi(d_1+k\gamma\sqrt{T-t}) - \Phi(d_2+k\gamma\sqrt{T-t})\right),
\end{aligned}
\end{equation*}
where
\begin{equation*}
\begin{aligned}
 d_1 = \frac{\log \left( \frac{\tilde{Z}_t}{a}\right) - \left( r+ \frac{\gamma^2}{2} \right)(T-t)}{\gamma\sqrt{T-t}}&& \text{and} && d_2 = \frac{\log \left( \frac{\tilde{Z}_t}{b}\right) - \left( r+ \frac{\gamma^2}{2} \right)(T-t)}{\gamma\sqrt{T-t}},
\end{aligned}
\end{equation*}
and $\Phi(\cdot)$ is the distribution function of the standard normal random variable.
\end{taggedlemma}
\begin{proof} Straightforward calculation.
\end{proof}

\begin{taggedlemma}{SL.2}[Solution to pointwise optimization problem]\label{lem:pointwise_optim_sol_linear_1}
Let $y\in(0, +\infty)$ be any fixed number. Then the expression
\begin{equation}\label{eq:pointwise_optim_sol_linear_1}
v^*:=v^*(y,\tilde{z}) = \left\{
\begin{aligned}
&I_{3}(y\tilde{z}), && \text{if }\, \tilde{z} \in (u_{M,3}'(\theta_3-)/y, u_{M,3}'(\theta_2+)/y), \\
&\theta_2, && \text{if }\, \tilde{z} \in [u_{M,3}'(\theta_2+)/y, u_{M,2}'(\theta_2-)/y], \\
&I_{2}(y\tilde{z}), && \text{if }\, \tilde{z} \in (u_{M,2}'(\theta_2-)/y, u_{M,2}'(\theta_1+)/y), \\
&\theta_1, && \text{if }\, \tilde{z} \in [u_{M,2}'(\theta_1+)/y, u_{M,1}'(\theta_1-)/y), \\
&\tilde{v}_1, && \text{if }\, \tilde{z} = u_{M,1}'(\theta_1-)/y, \\
&\theta_0, && \text{if }\, \tilde{z} \in (u_{M,1}'(\theta_1-)/y, +\infty),
\end{aligned}\right.
\end{equation}
solves for all $\tilde{z} \in (0, +\infty)$ the problem
\begin{equation}\label{eq:pointwise_OP}
 \max_{v \geq 0} \{u_M(v) - y\cdot \tilde{z}\cdot v \},
\end{equation}
where $u_M$ is defined in (\ref{eq:utility_piecewise}),  $I_i(\cdot):= (u_{M,i}')^{-1}(\cdot)$ for $i \in \{2,3\}$, $\tilde{v}_1$ is any number from  $[\theta_0, \theta_1]$.
\end{taggedlemma}
\begin{proof}[Proof of Lemma \ref{lem:pointwise_optim_sol_linear_1}]
The piecewise structure of the objective function \eqref{eq:pointwise_OP} motivates us to consider three subproblems, which are derived from the initial optimization problem by restricting the feasibility region to $[\theta_0,\theta_1]$, $[\theta_1,\theta_2]$ and $[\theta_2,\theta_3)$. Note that if we allow $\theta_0 < 0$, then the first subproblem will be restricted to $[0, \theta_1]$ due to the constraint $v \geq 0$ in (\ref{eq:point_optim_problem}). We derive optimal solutions of the subproblems depending on parameters $y \in (0,+\infty)$ and $\tilde{z} \in  (0, +\infty)$. Then for any fixed $y$ and $\tilde{z}$ we compare the solutions of the three subproblems to find the global optimizer of the initial problem (\ref{eq:point_optim_problem}). Finally, we write the global maximizer as a function of $y$ and $\tilde{z}$.

Consider the optimization problem:
\begin{equation}\tag{$P_1$}\label{eq:subproblem_1}
 \max_{v \in [\theta_0, \theta_1]}\{u_{M,1}(v) - y\tilde{z}v\}.
\end{equation}
According to \eqref{eq:utility_piecewise}, we $u_{M,1}(v)$ is linear and strictly increasing, whence we write $u_{M,1}(v) = a_1 v + b_1$ with $a_1 > 0$. Hence, the first subproblem is $\max_{v \in [0, \theta_1]}\{(a_1 v + b_1)- y\tilde{z}v\}$. Due to linearity of the objective function, we easily get the optimal solution depending on parameter values:
\begin{equation*}
 v_1^* = \left\{
 \begin{aligned}
 0,\qquad\,\, & \text{if }\, \tilde{z} > a_1/y;\\
 \tilde{v}_1\qquad\,\, & \text{if }\, \tilde{z} =  a_1/y;\\
 \theta_1, \qquad &\text{if }\, \tilde{z} < a_1/y,
 \end{aligned}
 \right.
\end{equation*}
where $\tilde{v}_1$ is any number from the interval $[0,\theta_1]$. So the optimum of this subproblem is unique for $\tilde{z} \neq a_1/y = u_{M,1}'(0+)/y = u_{M,1}'(\theta_1-)/y$.

Consider the second optimization subproblem:
\begin{equation}\tag{$P_2$}\label{eq:subproblem_2}
 \max_{v \in [\theta_1, \theta_2]} \{u_{M,2}(v)-y\tilde{z}v\} \Leftrightarrow  \min_{v \in [\theta_1, \theta_2]} \{-u_{M,2}(v)+y\tilde{z}v\}.
\end{equation}

The minimum exists because the objective function is continuous and the feasible set is compact. Constraints are linear and the objective function is strictly convex. The latter property follows from the fact that the function $-u_{M,2}(v)$ is strictly convex due to strict concavity of $u_{M,3}(v)$. The Slater's condition obviously holds. Therefore, the Karush-Kuhn-Tucker (KKT) conditions are necessary and sufficient for optimality. Due to strict convexity, the minimum is unique.

Using KKT conditions, we obtain the following optimal solution of the problem (\ref{eq:subproblem_2}) depending on the value of $y\tilde{z}$:
\begin{equation}\label{eq:w_2_optimal}
 v_2^*(y,\tilde{z})= \left\{
 \begin{aligned}
& \theta_1, &&\text{if }\, y\tilde{z} \geq u_{M,2}'(\theta_1+) \\ 
& I_2(y\tilde{z}), &&\text{if }\, u_{M,2}'(\theta_2-) < y\tilde{z} < u_{M,2}'(\theta_1+) \\
& \theta_2,&&\text{if }\, u_{M,2}'(\theta_2-) \geq y\tilde{z}
 \end{aligned}
 \right.
\end{equation}

Consider the optimization subproblem:
\begin{equation}\tag{$P_3$}\label{eq:subproblem_3}
 \max_{v \in [\theta_2, +\infty)} \{u_{M,3}(v)-y\tilde{z}v\} \Leftrightarrow  \min_{v \in [\theta_2, +\infty)} \{-u_{M,3}(v)+y\tilde{z}v\}.
\end{equation}
Analogously to the previous case, the optimal solution exists and is unique for any $y\tilde{z} > 0$. Using the corresponding KKT conditions, we obtain the optimal solution of (\ref{eq:subproblem_3}):
\begin{equation}\label{eq:w_3_optimal}
 v_3^* := v_3^*(y,\tilde{z}) = \left\{
 \begin{aligned}
 &\theta_2, &&\text{if }\, y\tilde{z} \geq u_{M,3}'(\theta_2+); \\ 
 &I_3(y\tilde{z}), &&\text{if }\, u_{M,3}'(\theta_2+) > y\tilde{z}.
 \end{aligned}
 \right.
\end{equation}

\begin{table}[!ht]
\begin{center}
  \begin{tabular}{@{} lccc @{}}
    \toprule
    $y\tilde{z} \in$ & $v_1^*$ & $v_2^*$ & $v_3^*$ \\
    \midrule
    $(0, u_{M,3}'(\theta_2+))$ & $\theta_1$ & $\theta_2$ & $I_3(y\tilde{z})$ \\
    $[u_{M,3}'(\theta_2+), u_{M,2}'(\theta_2-)]$ & $\theta_1$ & $\theta_2$ &  $\theta_2$ \\
    $(u_{M,2}'(\theta_2-), u_{M,2}'(\theta_1+))$ & $\theta_1$ & $I_2(y\tilde{z})$ & $\theta_2$ \\
    $[u_{M,2}'(\theta_1+), u_{M,1}'(\theta_1-))$ &$\theta_1$ & $\theta_1$ & $\theta_2$ \\
    $\{u_{M,1}'(\theta_1-) \}$ &  $\tilde{v}_1$ & $\theta_1$ & $\theta_2$ \\
    $(u_{M,1}'(\theta_1-), +\infty)$ & $0$ & $\theta_1$ & $\theta_2$ \\
    \bottomrule
  \end{tabular}
\end{center}
\caption{Optimal solutions of (\ref{eq:subproblem_1}), (\ref{eq:subproblem_2}), (\ref{eq:subproblem_3}) depending on the value of $y\tilde{z}$}
\label{tab:solutions_of_subprob_lin_1}
\end{table}

Now we show that the optimal solution of the problem $\max_{v \in [0;+\infty)}\{u_M(v) - y\tilde{z}v\}$ is given by (\ref{eq:pointwise_optim_sol_linear_1}).

Denote $f(v)=u_M(v)-y\tilde{z}v$ and $f_i(v)=u_{M,i}(v)-y\tilde{z}v,\, i \in \{1,2,3\}$. We distinguish between 6 cases, depending on the value of $y\tilde{z}$ and prove only the first two of them, as the rest are analogous.

1) If $\tilde{z} > u_{M,1}'(\theta_1-)/y = a_1/y$, then $v_1^* = 0$, $v_2^* = \theta_1$, $v_2^* = \theta_2$. We show now that $v_1^* = 0$ maximizes $f(v)$ on $v \geq 0$.

Take any $\bar{v} \in [0, \theta_1)$. Then:
\begin{equation*}
f(v_1^*) - f(\bar{v}) = f_1(v_1^*) - f_1(\bar{v}) \geq f_1(v_1^*) - \max_{v \in [0, \theta_1]} f_1(v) = f_1(v_1^*) - f_1(v_1^*) =  0
\end{equation*}
Take any $\bar{v} \in [\theta_1, \theta_2)$. Then:
\begin{flalign*}
f(v_1^*) - f(\bar{v}) & = f_1(v_1^*) - f_2(\bar{v}) \geq f_1(v_1^*) - \max_{v \in [\theta_1, \theta_2]} f_2(v) = f_1(v_1^*) - f_2(v_2^*) \\
& = f_1(v_1^*) - f_2(\theta_1) = f_1(v_1^*) - f_1(\theta_1) \geq f_1(v_1^*) - \max_{v \in [0, \theta_1]} f_1(v) =  f_1(v_1^*) - f_1(v_1^*) =  0. 
\end{flalign*}
We used optimal solutions of problems (\ref{eq:subproblem_1}) and (\ref{eq:subproblem_2}) for the corresponding values of $\tilde{z}$ as well as the fact that $f_1(\theta_1) = f_2(\theta_1)$.

Take any $\bar{v} \in [\theta_2, + \infty)$ and obtain:
\begin{flalign*}
 f(v_1^*) - f(\bar{v}) & = f_1(v_1^*) - f_3(\bar{v}) \geq f_1(v_1^*) - \max_{v \in [\theta_2, + \infty)} f_3(v) = f_1(v_1^*) - f_3(v_3^*) \\
& = f_1(v_1^*) - f_3(\theta_2) = f_1(v_1^*) - f_2(\theta_2)\geq f_1(v_1^*) - \max_{v \in [\theta_1, \theta_2]} f_2(v) \\
& = f_1(v_1^*) - f_1(\theta_1) \geq f_1(v_1^*) - \max_{v \in [0, \theta_1]} f_1(v) = f_1(v_1^*) - f_1(v_1^*) =  0.
\end{flalign*}
We used optimal solutions of problems (\ref{eq:subproblem_1}), (\ref{eq:subproblem_2}),  (\ref{eq:subproblem_3}) for the corresponding values of $\tilde{z}$ as well as the fact that $f_1(\theta_1) = f_2(\theta_1)$, $f_2(\theta_2) = f_3(\theta_2)$.

We conclude that for any $y>0$ and $\tilde{z} > u_{M,1}'(\theta_1-)/y$ the optimal solution of the problem $\max_{v \geq 0} \{u_M(v) - y\tilde{z}v \}$ is $0$.

2) If $\tilde{z} = u_{M,1}'(\theta_1-)/y = a_1/y$, then according to Table \ref{tab:solutions_of_subprob_lin_1} $v_1^* = \tilde{v}_1$, $v_2^* = \theta_1$, $v_2^* = \theta_2$, where $\tilde{v}_1 \in  [0, \theta_1]$. In this case $v_1^* = \tilde{v}_1$ solves $\max_{v\geq 0}f(v)$.
It can be concluded from the fact that $\tilde{z} = u_{M,1}'(\theta_1-)/y \geq u_{M,2}'(\theta_1+)/y > u_{M,2}'(\theta_2-)/y \geq  u_{M,3}'(\theta_2+)/y$ and the following relations
\begin{flalign*}
f(v_1^*) & = f_1(v^*) = f_1(\theta_1) = f_2(\theta_1) =  f_2(v_2^*) = \max_{v \in [\theta_1, \theta_2]} f_2(v) > f_2(\theta_2) = f_3(\theta_2) = f_3(v_3^*) = \max_{v \in [\theta_2, +\infty)} f_3(v). 
\end{flalign*}
First, we used that $f_1(v) = (a_1 v + b_1 ) - y(a_1/y) v = b_1 $ is constant. Then we used continuity of $f$ and the optimal solutions $v_2^*$ and $v_3^*$ of the corresponding subproblems for $\tilde{z} = u_{M,1}'(\theta_1-)/y$.

The remaining four cases are analogous to 1).
\end{proof}

\begin{taggedprop}{SP. 3}[Equations for computing $y^*$]\label{prop:equations_for_y}
Let $\theta_1$ and $s(\theta_1)$ be as defined in Corollary \ref{cor:fl3_optimal_terminal_vaule} depending on cases. Let  $\xi_1, d_1^A(\cdot), d_2^A(\cdot),d_1^B(\cdot), d_2^B(\cdot), d_3^B(\cdot), d_1^C(\cdot), d_2^C(\cdot), d_3^C(\cdot), d_4^C(\cdot)$ be as defined in Proposition \ref{prop:value_fct_manager}. Denote:
\begin{equation*}
  \begin{aligned}
    \xi_2 &= \ExpTwoPar{-}{0.5}.
  \end{aligned}
\end{equation*}
Then the explicit equation for computing the unique $y^*$ is given in\\
Case A: 
  \begin{equation*}
    \begin{aligned}
      & y^{-1/b_M}  \alpha^{(1-b_M)/b_M} \xi_1 \left(\Phi\left(d_1^A(y) + \left(1 - b_M^{-1}\right)\gamma \sqrt{T}\right) -  \Phi\left(d_2^A(y) + \left(1 - b_M^{-1}\right)\gamma \sqrt{T}\right) \right) \\
& + \left((1+m - \alpha^{-1} m)v_0 - \alpha^{-1} a_M \right) \xi_2 \left(\Phi\left(d_1^A(y) + \gamma \sqrt{T}\right) -  \Phi\left(d_2^A(y) + \gamma \sqrt{T}\right) \right) = v_0;
    \end{aligned}
  \end{equation*}
Case B:
   \begin{equation*}
    \begin{aligned}
      & y^{-1/b_M}  \alpha^{(1-b_M)/b_M} \xi_1 \left(\Phi\left(d_1^B(y) + \left(1 - b_M^{-1}\right)\gamma \sqrt{T}\right) -  \Phi\left(d_2^B(y) + \left(1 - b_M^{-1}\right)\gamma \sqrt{T}\right) \right) \\
& + \left((1+m - \alpha^{-1} m)v_0 - \alpha^{-1} a_M \right) \xi_2 \left(\Phi\left(d_1^B(y) + \gamma \sqrt{T}\right) -  \Phi\left(d_2^B(y) + \gamma \sqrt{T}\right) \right) \\
& + (1+m)v_0\xi_2 \left(\Phi\left(d_2^B(y) + \gamma \sqrt{T}\right) -  \Phi\left(d_3^B(y) + \gamma \sqrt{T}\right) \right) = v_0;
    \end{aligned}
  \end{equation*}
Case C:
 \begin{equation*}
    \begin{aligned}
     & y^{-1/b_M}  \alpha^{(1-b_M)/b_M} \xi_1 \left(\Phi\left(d_1^C(y) + \left(1 - b_M^{-1}\right)\gamma \sqrt{T}\right) -  \Phi\left(d_2^C(y) + \left(1 - b_M^{-1}\right)\gamma \sqrt{T}\right) \right) \\
& + \left((1+m - \alpha^{-1} m)v_0 - \alpha^{-1} a_M \right) \xi_2 \left(\Phi\left(d_1^C(y) + \gamma \sqrt{T}\right) -  \Phi\left(d_2^C(y) + \gamma \sqrt{T}\right) \right) \\
& + (1+m)v_0\xi_2 \left(\Phi\left(d_2^C(y) + \gamma \sqrt{T}\right) -  \Phi\left(d_3^C(y) + \gamma \sqrt{T}\right) \right) \\
& + y^{-1/b_M}\xi_1\PhiDifFin{d_3^C(y)}{d_4^C(y)}{+\left(1-b_M^{-1}\right)} \\
& + (v_0 - a_M)\xi_2 \PhiDifFin{d_3^C(y)}{d_4^C(y)}{+}{=} v_0.
    \end{aligned}
  \end{equation*}
\end{taggedprop}
\begin{proof}
According to Proposition \ref{th:optim_w_linear_part_1}, there exists a unique $y^*$ that solves  $\mathbb{E}\Bigl[\tilde{Z}_T  V_T^*\Bigr] = v_0$. Using Corollary \ref{cor:fl3_optimal_terminal_vaule} and the supplemetary Lemma \ref{lem:expectation_Zk_ab} from \ref{app:aux_proofs}, we obtain the explicit form of $\mathbb{E}\Bigl[\tilde{Z}_T V_T^*\Bigr]$. Since the calculations are straightforward but tedious, we do not provide them.
\end{proof}

\begin{taggedprop}{SP. 4}[First and second moment of the hedge-fund's optimal terminal value]\label{prop:expectation_of_V_T}
Let the manager's preferences be determined by $U_M$ as per \eqref{eq:utility_hara_manager_fl3}. Let $y^*$ be as defined in Corollary \ref{cor:fl3_optimal_terminal_vaule} and  $d_1^A(\cdot)$, $d_2^A(\cdot)$, $d_1^B(\cdot)$, $d_2^B(\cdot)$, $d_3^B(\cdot)$, $d_1^C(\cdot)$, $d_2^C(\cdot)$, $d_3^C(\cdot)$, $d_4^C(\cdot)$ be as defined in Proposition \ref{prop:value_fct_manager}. Denote:
\begin{equation*}
\begin{aligned}
\xi_3 &= \ExpTwoPar{b_M^{-1}}{0.5b_M^{-2}};\\
 \xi_4 &= \ExpTwoPar{2 b_M^{-1}}{2b_M^{-2}}.
\end{aligned}
\end{equation*}
The first two moments of the fund's optimal terminal value equal in\\
Case A:
     \begin{flalign*}
 \mathbb{E}\left[ V^*(T)\right]& = (y^*)^{-1/b_M}\alpha^{(1-b_M)/b_M} \xi_3 \left(\Phi\left(d_1^A -b_M^{-1}\gamma \sqrt{T}\right) - \Phi\left(d_2^A -b_M^{-1}\gamma \sqrt{T}\right) \right) & \\
 &\quad + \left((1+m - \alpha^{-1} m)v_0 - \alpha^{-1} a_M \right) \left(\Phi\left(d_1^A\right) -  \Phi\left(d_2^A\right) \right);\\
            \mathbb{E}\left[(V_T^*)^2\right] &= \alpha^{(2-2b_M)/b_M}(y^*)^{-2/b_M}\xi_4 \cdot \PhiDifFin{d_1^B}{d_2^B}{-2b_M^{-1}} \\
           &\quad+ 2 \alpha^{(1-b_M)/b_M}(y^*)^{-1/b_M}\left((1+m - \alpha^{-1} m)v_0 - \alpha^{-1} a_M \right)\xi_3\\
           &\quad \cdot \PhiDifFin{d_1^B}{d_2^B}{-b_M^{-1}}\\
           & \quad +  \left((1+m - \alpha^{-1} m)v_0 - \alpha^{-1} a_M \right)^2 \PhiDifNull{d_1^B}{d_2^B} + (1+m)^2 v_0^2\PhiDifNull{d_2^B}{d_3^B};
     \end{flalign*}
Case B:
     \begin{flalign*}
 \mathbb{E}\left[ V^*(T)\right] &= (y^*)^{-1/b_M}\alpha^{(1-b_M)/b_M} \xi_3 \left(\Phi\left(d_1^B (y^*) -b_M^{-1}\gamma \sqrt{T}\right) - \Phi\left(d_2^B (y^*) -b_M^{-1}\gamma \sqrt{T}\right) \right) &\\
 &\quad+ \left((1+m - \alpha^{-1} m)v_0 - \alpha^{-1} a_M \right) \left(\Phi\left(d_1^B (y^*)\right) -  \Phi\left(d_2^B (y^*)\right) \right)\\
 &\quad + (1+m)v_0\left(\Phi\left(d_2^B (y^*)\right) -  \Phi\left(d_3^B (y^*)\right) \right);\\
  \mathbb{E}\left[(V_T^*)^2\right] &= \alpha^{(2-2b_M)/b_M}(y^*)^{-2/b_M}\xi_4 \cdot \PhiDifFin{d_1^B (y^*)}{d_2^B (y^*)}{-2b_M^{-1}} \\
           &\quad+ 2 \alpha^{(1-b_M)/b_M}(y^*)^{-1/b_M}\left((1+m - \alpha^{-1} m) v_0 - \alpha^{-1} a_M \right)\xi_3\\
           &\quad \cdot \PhiDifFin{d_1^B (y^*)}{d_2^B (y^*)}{-b_M^{-1}}\\
           & \quad +  \left((1+m - \alpha^{-1} m)v_0 - \alpha^{-1} a_M \right)^2 \PhiDifNull{d_1^B (y^*)}{d_2^B (y^*)} \\
           &\quad+ (1+m)^2 v_0^2\PhiDifNull{d_2^B (y^*)}{d_3^B (y^*)};
      \end{flalign*}
Case C:
     \begin{flalign*}
 \mathbb{E}\left[ V^*(T)\right] &= (y^*)^{-1/b_M}\alpha^{(1-b_M)/b_M} \xi_3 \left(\Phi\left(d_1^C (y^*) -b_M^{-1}\gamma \sqrt{T}\right) - \Phi\left(d_2^C (y^*) -b_M^{-1}\gamma \sqrt{T}\right) \right) &\\
 &\quad+ \left((1+m - \alpha^{-1} m)v_0 - \alpha^{-1} a_M \right) \left(\Phi\left(d_1^C (y^*)\right) -  \Phi\left(d_2^C (y^*)\right) \right) \\
 &\quad + (1+m)v_0\left(\Phi\left(d_2^C (y^*)\right) -  \Phi\left(d_3^C (y^*)\right) \right) \\
 & \quad + (y^*)^{-1/b_M}\xi_3 \PhiDifFin{d_3^C (y^*)}{d_4^C (y^*)}{-b_M^{-1}} \\
 &\quad + (v_0 - a_M)\PhiDifNull{d_3^C (y^*)}{d_4^C (y^*)};\\
            \mathbb{E}\left[(V_T^*)^2\right] &= \alpha^{(2-2b_M)/b_M}(y^*)^{-2/b_M}\xi_4 \cdot \PhiDifFin{d_1^C(y^*)}{d_2^C(y^*)}{-2b_M^{-1}}\\
           & \quad+ 2 \alpha^{(1-b_M)/b_M}(y^*)^{-1/b_M}\left((1+m - \alpha^{-1} m)v_0 - \alpha^{-1} a_M \right)\xi_3\\
           & \quad \cdot \PhiDifFin{d_1^C(y^*)}{d_2^C(y^*)}{-b_M^{-1}}\\
           & \quad +  \left((1+m - \alpha^{-1} m)v_0 - \alpha^{-1} a_M \right)^2 \cdot \PhiDifNull{d_1^C(y^*)}{d_2^C(y^*)} \\
           & \quad + (1+m)^2 v_0^2\PhiDifNull{d_2^C(y^*)}{d_3^C(y^*)}\\
           &\quad + (y^*)^{-2/b_M}\xi_4\PhiDifFin{d_3^C(y^*)}{d_4^C(y^*)}{-2b_M^{-1}} +  2(y^*)^{-1/b_M}(v_0 - a_M)\xi_3\\
           &\quad \cdot \PhiDifFin{d_3^C(y^*)}{d_4^C(y^*)}{-b_M^{-1}}+ (v_0 - a_M)^2\PhiDifNull{d_3^C(y^*)}{d_4^C(y^*)}.
 \end{flalign*}
\end{taggedprop}
\begin{proof}
Straightforward calculation based on Corollary \ref{cor:fl3_optimal_terminal_vaule} and supplementary Lemma \ref{lem:expectation_Zk_ab}.
\end{proof}

\bibliography{Optimal_FL_Fee_Structures}

\end{document}